\documentclass[aps,pre,reprint,superscriptaddress,footinbib,floatfix]{revtex4-1}

\usepackage{amsmath}

\DeclareMathAlphabet{\mathsf}{OT1}{phv}{b}{n}

\newcommand{\crossVorg}{\ensuremath{%
         \setbox0=\hbox{$V$}
        V \kern-\wd0{\raise.3ex\hbox{$\relbar$}}}}

\newcommand{\crossVxx}[2]{%
	{\setbox0=\hbox{$#1#2V$}
         \setbox1=\hbox{$#1#2$}
         \setbox2=\hbox{$#1V$}
         \dimen1=\wd0
	 \advance\dimen1-\wd1
         \raise.2\ht0\hbox{$#1#2$}\kern-.4\wd0}}

\usepackage{pifont}

\usepackage{multirow}
\usepackage{setspace}
\usepackage{enumitem}
\usepackage{verbatim}

\usepackage{amssymb}
\usepackage{amsbsy}
\usepackage[T1]{fontenc}
\usepackage{amsmath}
\usepackage{mathtools}

\usepackage{graphicx}
\usepackage{xcolor}
\usepackage{stmaryrd}
\usepackage{mathrsfs}
\usepackage[mathscr]{euscript}
\usepackage{tikz}
\usetikzlibrary{shapes}
\usepackage{cancel}
\usepackage[normalem]{ulem}

\usepackage[colorinlistoftodos]{todonotes}
\usepackage{verbatim}
\usepackage{latexsym}
\usepackage{textcomp}
\usepackage{bm}
\usepackage[Euler]{upgreek}
\usepackage{longtable}
\usepackage{subfigure}

\usepackage[colorlinks,allcolors=cyan!70!black]{hyperref}

\usepackage{epsfig}
\usepackage{dcolumn}

\usepackage{booktabs}

\newcommand{\qed}{\nobreak \ifvmode \relax \else
      \ifdim\lastskip<1.5em \hskip-\lastskip
      \hskip1.5em plus0em minus0.5em \fi \nobreak
      \vrule height0.75em width0.5em depth0.25em\fi}

\newcommand{\la}{\ensuremath{\langle}}
\newcommand{\ra}{\ensuremath{\rangle}}

\newcommand{\angstrom}{\mbox{\normalfont\AA}}

\DeclareMathAlphabet\mathbfcal{OMS}{cmsy}{b}{n}
\DeclareMathAlphabet{\mathbfsf}{\encodingdefault}{\sfdefault}{bx}{sl}

\usepackage[colorlinks,allcolors=cyan!70!black]{hyperref}

\begin{document}


\title{Universality of dilute solutions of ring polymers in the thermal crossover region between $\theta$ and athermal solvents}


\author{Aritra Santra}
\affiliation{Department of Chemical Engineering, Monash University,
Melbourne, VIC 3800, Australia}
\author{J. Ravi Prakash}
\email{ravi.jagadeeshan@monash.edu}
\affiliation{Department of Chemical Engineering, Monash University,
Melbourne, VIC 3800, Australia}
 \homepage{https://users.monash.edu.au/~rprakash/}
 

\date{\today}



\begin{abstract}
Due to their unique topology of having no chain ends, dilute solutions of ring polymers exhibit behaviour distinct from their linear chain counterparts. The universality of their static and dynamic properties, as a function of solvent quality $z$ in the thermal crossover regime between $\theta$ and athermal solvents, is studied here using Brownian dynamics simulations. The universal ratio $U_{\text{RD}}$ of the radius of gyration $R_g$ to the hydrodynamic radius $R_H$ is determined, and a comparative study of the swelling ratio $\alpha_g$ of the radius of gyration, the swelling ratio $\alpha_H$ of the hydrodynamic radius, and the swelling ratio $\alpha_X$ of the mean polymer stretch $X$ along the $x$-axis, for linear and ring polymers, is carried out. The ratio $U_{\text{RD}}$ for dilute ring polymer solutions is found to converge asymptotically to a constant value as $z \to \infty$, which is a major difference from the behaviour of solutions of linear chains, where no such asymptotic limit exists. Additionally, the ratio of the mean stretch along the $x$-axis to the hydrodynamic radius, $(X/R_H)$, is found to be independent of $z$ for polymeric rings, unlike in the case for linear polymers. These results indicate a fundamental difference in the scaling of static and dynamic properties of rings and linear chains in the thermal crossover regime.

\end{abstract}

\maketitle


\section{\label{sec:intro} Introduction}

The universal behaviour of various static and dynamic properties of dilute solutions of linear homopolymers,  such as the radius of gyration, or the hydrodynamic and viscometric radii, is well known~\cite{deGennes79,RubColby2003,Prakash2019}. For instance, in the limit of high molecular weight, both the radius of gyration and hydrodynamic radius exhibit power law scaling with molecular weight under $\theta$ or athermal solvent conditions. Many numerical studies have investigated this behaviour~\cite{Baschnagel2004,Hsu2011,Duenweg2018,Prakash2019}, and recently high precision Monte Carlo simulation algorithms have been developed to estimate the values of these power law exponents very accurately~\cite{Clisby2010,Clisby2016}. The self-similar fractal structure of polymer chains is the key reason for observing such universality~\cite{Duenweg2018}. While power law scaling is typically observed for high molecular weight polymers, linear homopolymer solutions at temperatures between the $\theta$ and athermal limits also exhibit universality when their behaviour is interpreted in terms of the solvent quality parameter, $z$, which is a function of both the solution temperature $T$ and the molecular weight $M$, and is defined as $z=k_0\,(1- {\Theta}/T)\sqrt{M}$. Here ${\Theta}$ is the temperature at which $\theta$-solvent conditions are observed and $k_0$ is a chemistry-dependent constant, with units of $M^{-1/2}$, since $z$ is dimensionless~\cite{Miyaki1980,Fujita,Hayward,Rondelez,Schafer,Yamakawa1971}. The limit $z=0$ indicates a $\theta$-solvent, while $z\rightarrow \infty$ corresponds to the athermal solvent limit. The commonly used Flory $\chi$-parameter can be related to $z$ by comparing the terms involving the second virial coefficient when the two definitions of solvent quality are used in the virial expansion for the osmotic pressure. It can be shown that $1-2\chi = 1- {\Theta}/T = z/(k_0\sqrt{M})$~\cite{RubColby2003}. The thermal crossover regime corresponds to values of $z$ that lie between the $\theta$ and athermal solvent limits, i.e., in the range $0\le z<\infty$. In the crossover regime, the universal behaviour of polymer solutions is typically expressed in terms of the ratios of different static and dynamic properties, or by swelling ratios, which are defined as the ratio of a property at a temperature $T> \Theta$ to its value at the $\theta$-temperature. Several experimental studies and molecular simulations of dilute linear homopolymer solutions have demonstrated the collapse of data for the various ratios on to master curves when interpreted in terms of the solvent quality $z$~\cite{Miyaki1980,Fujita,Rondelez,Hayward,Tominaga2002,Sharad2014,Schafer,Kroger2000,KumarRavi,Sunthar2006}.  For polystyrene in cyclohexane, \citet{KumarRavi} found the value $k_0 = 0.0063$~(g/mol)$^{-1/2}$ collapsed the data of \citet{Fujita} onto the universal swelling curve for the radius of gyration predicted by BD simulations, while \citet{Pan2014a} determined that a value of $k_0$ to be $0.0047 \pm 0.0003$~(g/mol)$^{-1/2}$ collapsed data for linear DNA molecules in the range of $3-300$ kbp onto the universal swelling curve for the hydrodynamic radius, predicted by BD simulations. Apart from linear homopolymers, there have also been several experimental and theoretical studies that have been carried out to understand the nature of universality and the effect of solvent quality on polymers with more complex intramolecular interactions and topologies~\cite{Douglas84,freireetal84,lipsonetal87,ohnobinder88,batoulis89,douglasetal1990,Mourey1992,okumoto99,striolo2000,Tande2001,Likos2004,PMaiti2007,Bosko2011,Aritra2019}. In contrast, however, studies examining the universal behaviour of dilute ring polymer solutions in the intermediate regime between $\theta$ and athermal solvents have been relatively sparse, and in particular, to our knowledge, there are no experimental or computational studies that systematically examine their universal behaviour in terms of the solvent quality parameter $z$. The aim of the present work is to study the universal thermal crossover behaviour of dilute ring polymer solutions with the help of Brownian dynamics simulations.

While the thermal crossover of ring polymer solutions has not been studied in great detail, there have nevertheless been several studies that have examined the differences in the properties of ring and linear polymer solutions, which have provided a great deal of insight into the role that topology plays in the determination of solution properties. The key results of these studies, which are relevant to the current work, are summarised point-wise below.

\begin{enumerate}[itemsep=3pt, topsep=2pt]

\item[(i)] Undoubtedly, the most significant finding of experimental~\cite{Roovers1985,Lutz1986,Kovacs1987,Takano2009,Vlassopoulos2015}, analytical~\cite{descloizeaux1981,Grosberg2000}, and computational~\cite{Jang2003,Narros2013,Li2016} studies is that the $\theta$-temperature for solutions of rings, $\Theta_R$, is less than that for linear polymer solutions, $\Theta_L$, in identical solvents. In other words, at a given temperature, a ring polymer would experience effectively better solvent conditions than a linear chain with the same molecular weight. For instance, the $\theta$-temperature for a solution of polystyrene rings with weight-average molar mass $M_w = 161 000$ g/mol in $d_{12}$-cyclohexane was estimated to be $31.5 \pm 1^\circ$C, whereas a value of 38.3$^\circ$C was found for its linear counterpart~\cite{Vlassopoulos2015}. Cyclization consequently leads in this case to the lowering of the $\theta$-temperature by about 6.8$^\circ$C. 

\item[(ii)] It follows, not unexpectedly, as shown by several studies~\cite{Roovers1985,Takano2009,Vlassopoulos2015,descloizeaux1981,Deguchi1997,Narros2013,Li2016,Gartner2019} that the second virial coefficient $B_2$ for ring polymer solutions is positive at $\Theta_L$. Experiments carried out  at $\Theta_L$ by \citet{Takano2009} reveal that $B_2$ scales as $M^{-0.34}$, confirming the earlier analytical prediction of \citet{Deguchi1997}. By careful off-lattice Monte Carlo simulations of a bead-spring chain model for rings, \citet{Li2016} have established that  at $\Theta_L$, the second virial coefficient for rings with small bead-numbers scales with the number of beads $N$ as $B_2 \sim N^{-1/2}$ due to 3-body interactions (as in the case of linear chains), but crosses over to $N^{-0.34}$ for larger rings. The improved solvent quality for ring polymer solutions is consequently due to both 3-body interactions and to topological constraints.

\item[(iii)] At the $\theta$-temperature for rings,  $\Theta_R$, analytical results~\cite{Kramers1946,Zimm1949,Yamakawa1971}, computer simulations~\cite{Suzuki2011,Narros2013} and experiments~\cite{Kovacs1987,Vlassopoulos2015} have established that rings behave like ideal Gaussian chains and obey random-walk statistics, with the radius of gyration obeying $R_g \sim M^{1/2}$. At $\Theta_L$, however, the experiments of \citet{Takano2012} suggest that the radius of gyration of rings scales with molecular weight with an \textit{effective} Flory exponent, $\nu_\text{eff} =0.53$, which has also been observed in explicit solvent molecular dynamics simulations by \citet{Gartner2019}. In athermal solvents, on the other hand, the differences between rings and linear chains in the scaling of the radius gyration with molecular weight vanishes, and the Flory exponent becomes similar in both solutions~\cite{Takano2012,Vlassopoulos2015,Gartner2019,Zifferer2001,Suzuki2011,Narros2013}.

\item[(iv)] The most commonly measured universal ratio in ring polymer solutions is the so-called \textit{g-factor}, which is the square of the ratio of the radius of gyration of ring polymers to that of linear chains, $R_{gR}^2/R_{gL}^2$. For Gaussian rings, its value has been shown analytically to be 0.5 by \citet{Zimm1949}. It should be noted that to validate this prediction experimentally, $R_{g}$ for rings and linear chains would have to be measured at the respective $\theta$-temperatures for the two solutions, since these differ for the two topologies. This has been done carefully by \citet{Roovers1985} and \citet{Kovacs1987}, who obtain good agreement with the analytical prediction. Indeed, if one uses the measurement at $\Theta_R$ by \citet{Vlassopoulos2015} of $R_{gR}^\theta = 76 \, \angstrom$ for $161 000$ g/mol molecular weight polystyrene rings in $d_{12}$-cyclohexane, and calculates the radius of gyration for a linear chain with the same molecular weight in the same solvent using the expression given by \citet{Takano2012}, $R_{gL}^\theta = 0.0315 M_w^{0.5}$, one obtains $R_{gL}^\theta =  107.3 \, \angstrom$, and consequently, $\left( R^\theta_{gR} /R^\theta_{gL} \right)^2 =0.5$. Measurements of this ratio at $\Theta_L$ will clearly be $> 0.5$ since rings are still swollen at this temperature, and can be expected to be a function of molecular weight with the scaling, $R_{gR}^2/R_{gL}^2 \sim M^{2\nu_\text{eff}  - 1}$. This was indeed reported by \citet{Takano2012} who found values for the \textit{g-factor} ranging from 0.557 to 0.73 for molecular weights between 17,000 and 570,000  g/mol. 

\item[(v)] While in general the value of the ratio $R_{gR}^2/R_{gL}^2$ in good solvents is higher than it is in $\theta$-solvents, the situation is more complicated, with a wide range of values reported depending on the methodology used for the estimation of the ratio. For instance, analytical calculations by \citet{Prentis1982} suggest a value of 0.568, while those by \citet{Douglas84} give 0.516. Computer simulations by \citet{Zifferer2001} lead to a value of 0.536, while a value of 0.56 can be extracted from the Monte Carlo simulation results published by \citet{Suzuki2011}, with a similar value reported by \citet{Gartner2019}. Experimental measurements by \citet{Higgins1979} and \citet{Lutz1986} lead to a value of 0.53, while \citet{Ragnetti85} report a measured value of 0.56. As discussed in greater detail subsequently, based on the measurements reported in \citet{Vlassopoulos2015}, one can estimate that their results lead to a value of the ratio $R_{gR}^2/R_{gL}^2 = 0.55$. 

\end{enumerate}

There are several other properties of ring polymer solutions that have been reported in the literature (some of which will be discussed subsequently), however, the points above are a broad summary of the behaviour of dilute ring polymer solutions that is relevant to the current work, which is focussed on examining the universal behaviour of static and dynamic properties of these solutions in the thermal crossover regime with the help of Brownian dynamics simulations. In particular, the radius of gyration $R_g$ and the mean-stretch along the $x$-axis $X$, which are static properties, and the diffusivity $D$ (or equivalently the hydrodynamic radius $R_H$), which are dynamic properties have been investigated in terms of various ratios of these properties, for both rings and linear polymers. A coarse-grained bead-spring chain model has been used to represent linear and ring topologies, and both excluded volume (EV) and hydrodynamic interactions (HI) have been incorporated since they are essential for obtaining an accurate description of the static and dynamic behaviour of dilute polymer solutions, respectively~\cite{Prakash2019}.

Within the framework of molecular simulations, it is common to account for the presence of excluded volume interactions with a Lennard-Jones (LJ) or an equivalent pair-wise potential between beads, in which the degree of effective attraction or repulsion between monomers is controlled by varying the magnitude of the attractive well depth of the potential, $\epsilon_\text{LJ}$. Whilst simulating athermal solvents is straightforward with the LJ or equivalent potentials, the exploration of the crossover region requires an elaborate procedure for mapping values of $\epsilon_\text{LJ}$ onto the solvent quality $z$~\cite{graetal999,Aritra2019}. The situation becomes even more involved when it is desired to carry out comparative simulations of rings and linear polymer solutions. As pointed by \citet{Gartner2019}, who performed explicit solvent molecular dynamics simulations for this reason, using the same EV potential parameters in the two solutions does not imply their solvent quality is the same, since their $\theta$-temperatures are different. Identical solvent quality requires that solutions of rings and linear chains with identical molecular weights are at different temperatures such that their distance from their respective $\theta$-temperatures are identical, i.e., $(1- {\Theta}/T)$ is the same in both. All these issues can be circumvented with the use of a much more convenient representation of the excluded volume energy as given by the purely repulsive narrow Gaussian potential~\cite{RaviExclu,Ottinger1996}. With this potential, which is described in greater detail subsequently in Section~\ref{sec:Model_rings}, simulations can be carried out (in both kinds of solutions) directly at desired values of the solvent quality $z$. In particular, $\theta$-solvent conditions are obtained by switching off EV interactions entirely, with the bead-spring chains treated as phantom chains obeying random-walk statistics. The narrow Gaussian potential has been used extensively in Brownian dynamics simulations of linear polymers and dendrimers, and has proven to be extremely useful for examining the role of solvent quality in determining solution behaviour at equilibrium and in flow, and for establishing the existence of universal behaviour~\cite{RaviExclu,Jendrejack20027752,KumarRavi,Sunthar2005,Sunthar2006,Bosko2011,AJainPRL,Sharad2014,Saadat2015,Sasmal2017}. Here, the narrow Gaussian potential is used to predict the properties of ring and linear polymer solutions at the same values of the solvent quality $z$. Comparison of experimental measurements of ring and linear polymer solutions with simulation results under identical solvent conditions would require that appropriate polymer molecular weights and temperatures of the two solutions are considered such that, $z = k_0\,(1- {\Theta_R}/T_R) \sqrt{M_R} = k_0\,(1- {\Theta_L}/T_L) \sqrt{M_L}$, where $T_R, T_L$ and $M_R, M_L$ refer to the ring and linear solution temperatures and molecular weights, respectively.

The incorporation of fluctuating hydrodynamic interactions in simulations, which have been implemented here with the Rotne-Prager-Yamakawa (RPY) tensor (as described in greater detail in Section~\ref{sec:Model_rings}), is essential for obtaining accurate predictions of dynamic properties~\cite{Prakash2019}. An important difference in the present work from earlier studies on ring polymers is the method used for the calculation of the diffusivity, which is required for the calculation of $R_H$. While most previous simulations compute the diffusivity of polymer chains of different topology using Kirkwood's expression~\cite{Kirkwood1954,Uehara2014,Uehara2016}, here it is shown that in order to evaluate the long-time diffusivity accurately it is necessary to apply Fixman's correction factor to Kirkwood's expression, which is derived from a dynamic correlation function~\cite{Fixman1981}. Simulations carried out by \citet{Liu2003} and \citet{Sunthar2006} have shown that for linear polymers, Kirkwood's expression corrected with Fixman's factor is essential for an accurate evaluation of the long-time diffusion coefficient. Kirkwood's expression leads to the short-time diffusivity, and while the difference without Fixman's correction is small for relatively short finite chain lengths, it has significant implications for the prediction of the universal swelling of the hydrodynamic radius, which is obtained in the infinite chain length limit~\cite{Sunthar2006}. 

Prediction of solution properties in the infinite chain length limit are found by carrying out simulations at finite chain lengths and extrapolating to the number of beads in the chain going to infinity. The universal nature of the results obtained in this manner is established by demonstrating the independence of the long chain limit from the choice of model parameters. As a consequence, the predictions of the various universal ratios in terms of solvent quality presented here are readily comparable with experiments since they are independent of any model parameters.

The paper is organised as follows. In Section~\ref{sec:Model_rings} the modelling and simulation methodology used for describing ring polymers is discussed. Evaluation of universal ratios involving static properties, such as the radius of gyration $R_g$ and the mean-stretch along the $x$-axis $X$, is described in Section~\ref{sec:Uniratio_rings}. In Section~\ref{sec:UniratioHI_rings}, the methods for computing the diffusivity $D$ and the hydrodynamic radius $R_H$ are  presented. Different universal ratios involving $R_g, R_H$ and $X$ are also discussed in this section. Finally, the key conclusions are summarised in the Section~\ref{RingConclusions}.

\section{Modelling and simulations}\label{sec:Model_rings}

A polymeric ring is modelled as a bead-spring chain with no chain ends, with adjacent beads connected by a finitely extensible non-linear elastic (FENE) spring, whose potential energy $U_\text{FENE}$ is given by the expression,
\begin{equation}\label{Eq:Ufene_dim}
U_\text{FENE} (Q) = -\frac{1}{2}\,H{Q_0}^2\,\ln\left(1-\frac{{Q}^2}{{Q_0}^2}\right)
\end{equation}
where $H$ is the spring constant, $Q_0$ is the maximum stretchable length of the spring and $Q$ is the instantaneous stretch of the spring. The time evolution of the position of a bead $\mu$, $\bm{r}_{\mu}(t)$, with excluded volume and hydrodynamic interactions implemented, can be described by an It\^{o} stochastic differential equation~\cite{Ottinger1996,PraRavi04,Stoltz2006}, 
\begin{widetext}
\begin{align}\label{GovEqn}
\begin{aligned}
\bm{r}^*_\mu(t^* + \Delta t^*) = \bm{r}^*_\mu(t^*) + \left(\bm{\kappa}^*\cdot\textbf{r}^*_{\nu}(t^*)\right)\Delta t^* + \frac{\Delta t^*}{4} \sum\limits_{\nu=1}^N\mathbf D^*_{\mu\nu}\cdot(\mathbf F_\nu^{*s}+ \mathbf F_\nu^{*\textrm{EV}}) +\frac{1}{\sqrt{2}}\sum\limits_{\nu=1}^N \mathbf B^*_{\mu\nu}\cdot\Delta \mathbf W^*_\nu
\end{aligned}
\end{align}
\end{widetext}
where the superscript  ($\ast$) denotes a non-dimensional quantity. Length and time scales have been non-dimensionalised with $l_H=\sqrt{k_BT/H}$ and $\lambda_H=\zeta/4H$, respectively, (i.e., $\bm{r}^*_\mu = \bm{r}_\mu/l_H$ and $t^* = t/\lambda_H$, etc), where $k_B$ is the Boltzmann constant, and $\zeta=6\pi\eta_s a$ is the Stokes friction coefficient of a spherical bead of radius $a$, with $\eta_s$ being the solvent viscosity. The quantity $\bm{\kappa}^* = (\bm{\nabla^* v^*})^{T}$ is a $3\times 3$ tensor, where $\bm{v}^*$ represents the unperturbed non-dimensional solvent velocity field. This term is considered to be zero in absence of any external flow field. $\Delta\mathbf W^*_\nu$ is a non-dimensional Wiener process, whose components are obtained from a real-valued Gaussian distribution with zero mean and variance $\Delta t^*$. $\mathbf{B}^*_{\mu\nu }$ is a non-dimensional tensor whose evaluation requires the decomposition of the diffusion tensor $\mathbf D^*_{\mu\nu}$, defined as $\mathbf D^*_{\mu\nu} = \delta_{\mu\nu} \pmb \delta + \pmb \Omega^*_{\mu\nu}$, where $\delta_{\mu\nu}$ is the Kronecker delta, $\pmb \delta$ is the unit tensor, and $\pmb{\Omega}^*_{\mu\nu}$ is the hydrodynamic interaction tensor. Block matrices $\mathcal{D}$ and $\mathcal{B}$ consisting of $N \times N$ blocks each having dimensions of $3 \times 3$ are defined such that the $(\mu,\nu)$-th block of $\mathcal{D}$ contains the components of the diffusion tensor $\mathbf{D}^*_{\mu\nu }$, whereas, the corresponding block of $\mathcal{B}$ is equal to $\mathbf{B}^*_{ \mu\nu}$. The decomposition rule for obtaining $\mathcal{B}$ can be expressed as $\mathcal{B} \cdot {\mathcal{B}}^{\textsc{t}} = \mathcal{D}\label{decomp}$. In the present study, the regularized Rotne-Prager-Yamakawa (RPY) tensor is used to compute hydrodynamic interactions,
\begin{equation}
{\pmb{\Omega}^*_{\mu \nu}} = {\pmb{\Omega}^*} ( {\bm{r}^*_{\mu}} - {\bm{r}^*_{\nu}} )
\end{equation}
where 
\begin{equation}
\pmb{\Omega}^*(\bm{r}^*) =  {\Omega_1{ \pmb \delta} +\Omega_2\frac{\bm{r^* r^*}}{{r}^{*2}}} \, ; \,\, \text{with} \, \left\vert \bm{r}^* \right\vert = r^*
\end{equation}
and
\begin{equation*}
\Omega_1 = \begin{cases} \dfrac{3\sqrt{\pi}}{4} \dfrac{h^*}{r^*}\left({1+\dfrac{2\pi}{3}\dfrac{{h^*}^2}{{r}^{*2}}}\right) & \text{for} \quad r^*\ge2\sqrt{\pi}h^* \\
 1- \dfrac{9}{32} \dfrac{r^*}{h^*\sqrt{\pi}} & \text{for} \quad r^*\leq 2\sqrt{\pi}h^* 
\end{cases}
\end{equation*}
\vspace{5pt}
\begin{equation*}
\Omega_2 = \begin{cases} \dfrac{3\sqrt{\pi}}{4} \dfrac{h^*}{r^*} \left({1-\dfrac{2\pi}{3}\dfrac{{h^*}^2}{{r}^{*2}}}\right) & \text{for} \quad r^*\ge2\sqrt{\pi}h^* \\
 \dfrac{3}{32} \dfrac{r^*}{h^*\sqrt{\pi}} & \text{for} \quad r^*\leq 2\sqrt{\pi}h^* 
\end{cases}
\end{equation*}
The hydrodynamic interaction parameter $h^* = a/(\sqrt{\pi k_BT/H})$ is the dimensionless bead radius. The diffusion tensor is decomposed using the Fixman's Chebyshev polynomial approximation which has been widely used earlier for linear polymers in both single chain~\cite{fixman1986implicit,PraRavi04,PrabhakarSFG,jendrejack,Kroger2000} and multi-chain BD simulations~\cite{JainPRE2012,Stoltz2006,SaadatSemi,JainSasmal2015}. The quantity $\bm{F}_\nu^{*S}$ is the net dimensionless spring force on the $\nu$-th bead, $\bm{F}_\nu^{*S} = \bm{F}_\nu^{*c} - \bm{F}_{\nu-1}^{*c}$, where $\bm{F}^{*c}_\nu = \partial U^*_\text{FENE}/ \partial \bm{Q}^*_\nu$ is the non-dimensional tension in the spring connecting the $(\nu+1)$ and the $\nu$-th beads, with $U^*_\text{FENE} =  U_\text{FENE}/ k_BT$ representing the non-dimensional potential energy, and $\bm{Q}^*_\nu$ being the non-dimensional connector vector between the beads, $\bm{Q}^*_\nu = \bm{r}^*_{\nu+1} - \bm{r}^*_\nu$, for $\nu = 1,\ldots,N-1$. From Eq.~(\ref{Eq:Ufene_dim}) it follows that,
\begin{align}
\bm{F}^{*c}_\nu= \frac{\bm{Q}^*_\nu}{1 - \bm{Q}^{*2}_\nu/b}
\label{eq:FENE}
\end{align}
where $b={Q_0}^2/l_H^2$ is the well known FENE $b$-parameter~\cite{Bird1987,PraRavi04}. 
Unless mentioned otherwise, all the simulations reported here are carried out with $b$ equal to $50.0$. Note that such a large value of $b$ raises  the possibility of self-crossing of the chain. This would pose a problem when modelling  the dynamics of dense entangled systems. However, since the focus here is on polymer solution statics and dynamics in the very dilute unentangled regime, topological constraints do not play a role. Indeed, self-crossings are expected to speed up the exploration of phase space and are advantageous for our purposes. The non-dimensional excluded volume potential energy, $E^* = E/k_B T$, is found by summing the interaction energy over all pairs of beads $\mu$ and $\nu$,
\begin{equation*}
E^*=\frac{1}{2} \sum_{\substack{\mu, v=1 \\ \mu \neq v}}^{N_b} E_{\nu\mu}^*\left(\bm{r}^*_{v}-\bm{r}^*_{\mu}\right)
\end{equation*}
where $E_{\nu\mu}^*\left(\bm{r}^*_{v}-\bm{r}^*_{\mu}\right)$ is a short-range function which is assumed here to be given by a narrow Gaussian potential~\cite{Ottinger1996,RaviExclu},
\begin{equation}
 E_{\nu\mu}^*(\bm{r}^*_{\nu\mu}) = \left(\frac{z^*}{{d^*}^3}\right) \exp{\left\lbrace -\frac{1}{2}\frac{\bm{r}_{\nu\mu}^{*2}}{{d^*}^2}\right\rbrace}
\end{equation}
where $\bm{r}^*_{\nu\mu} = \bm{r}^*_{\nu}-\bm{r}^*_{\mu}$, and the parameters $z^*$ and $d^*$ are non-dimensional quantities which characterise the narrow Gaussian potential: $z^*$ measures the strength of the excluded volume interaction, while $d^*$ is a measure of the range  of  excluded  volume  interaction.  The  narrow  Gaussian  potential  is  a  means  of regularizing the Dirac delta potential since it reduces to a $\delta$-function potential in the limit of $d^*$ tending to zero. The contribution of the non-dimensional force due to excluded volume interactions, $\mathbf F_\nu^{*\textrm{EV}}$, on bead $\nu$ is then,
\begin{equation}
 \mathbf F_\nu^{*\textrm{EV}} = -\frac{\partial E^*}{\partial \bm{r}^*_{\nu}} = - \sum_{\substack{\mu=1 \\ \mu\ne\nu}}^{N_b}\frac{\partial}{\partial\bm{r}^*_{\nu}}E_{\nu\mu}^*(\bm{r}^*_{\nu\mu})
\end{equation}
The solvent quality is essentially a function of the temperature and chain length, and in the context of the narrow Gaussian potential, it is defined as $z=z^*\sqrt{N_b}$, with $z^*$ representing the distance of the solution temperature from the $\theta$-temperature. As discussed earlier, this enables the simulation of both linear and ring polymer solutions at the same solvent quality, even though their $\theta$-temperatures may be different. As is common in most molecular simulations, the $\theta$-condition is replicated here by switching off excluded volume interactions altogether~\cite{Deguchi14JCP,Uehara2016,Sunthar2005}. 

A Brownian dynamics (BD) simulation algorithm has been implemented with an implicit predictor-corrector algorithm to solve the It\^{o} stochastic differential equation, Eq.~(\ref{GovEqn}), similar to that suggested by \"{O}ttinger~\citep{Ottinger1996} and implemented in their study by Prabhakar et al.~\citep{PraRavi04}. The simulations are carried out over a range of solvent qualities by varying the parameter $z^*$ and the chain length $N_b$. The typical values of chain length used in the simulations range from $N_b=10$ to $80$. The value of the parameter $z^* = z/\sqrt{N_b}$ depends on both $z$ and $N_b$. Here, we have used values of $z$ ranging from $0$ to $5$. Unless mentioned otherwise, all the simulations are carried out with the value of the range of interaction parameter $d^*$ set equal to $1.0$. Dynamic properties of ring and linear polymers are computed with hydrodynamic interactions, and the value  $h^*=0.24$ is used in all such simulations, unless stated otherwise. A typical simulation for a single chain ring polymer consists of running the simulation for about 5 to 10 non-dimensional Rouse relaxation times for linear chains ($\tau_R^*$) in the equilibration step followed by 5 non-dimensional Rouse relaxation times for the production run, where the non-dimensional Rouse relaxation time for linear chains is defined by~\cite{Bird1987},
\begin{equation}
 \tau_R^* = \frac{1}{2\sin^2 ({\pi}/{2N_b})} 
\end{equation} 

All the simulations are carried out with a non-dimensional time step of $\Delta t^* = 0.001$ and the output data are collected at an interval of $0.5$ non-dimensional time units. The static equilibrium properties are computed as a block ensemble average over several trajectories. Time averages for different properties are calculated at first for each independent trajectory, followed by calculation of ensemble averages over such independent time averages. The error of the mean is estimated over the ensemble of independent time averages. On the other hand, while computing dynamic properties such as the long-time diffusivity, the production run typically consists of several thousand non-dimensional time units, and in this case, properties are simply calculated based on ensemble averages over several independent trajectories. Typically, $500$ to $1000$ independent trajectories are used to calculate different static and dynamic properties.

\begin{figure*}[th]
  \begin{center}
		\begin{tabular}{cc}
		 \hspace{-1cm}
			\resizebox{9cm}{!}{\includegraphics*[width=4cm]{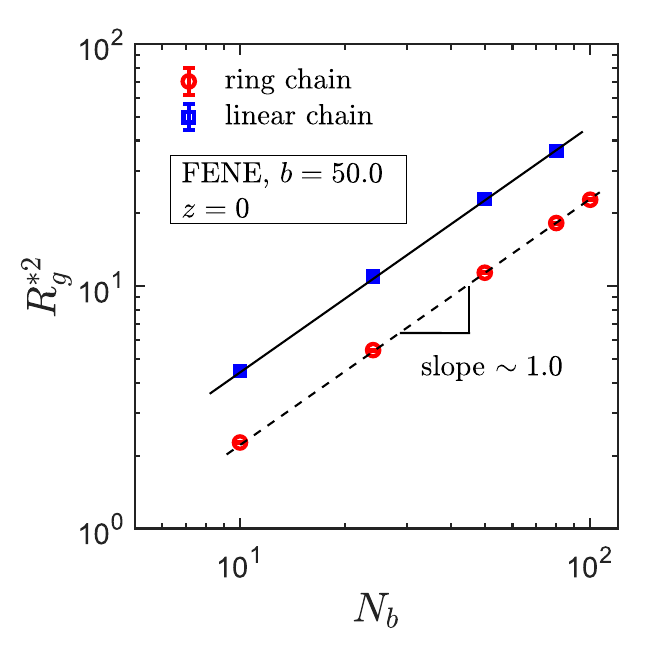}}   &
		 \hspace{-0.5cm} 	
			\resizebox{9.3cm}{!} {\includegraphics*[width=4cm]{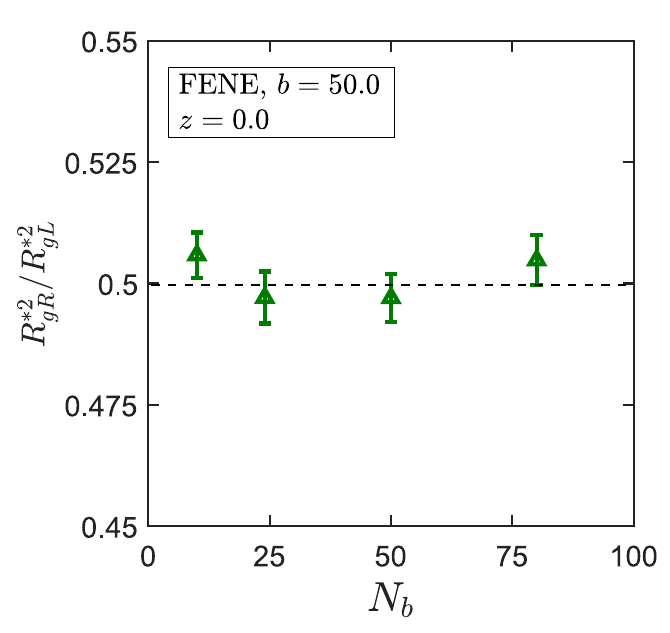}} \\
			(a) & (b) \\
		\end{tabular}
	\end{center}
	\vskip-15pt
\caption{(a) Mean square radius of gyration, $R_g^{*2}$, as function of chain length, $N_b$, for ring and linear polymers. (b) The variation of the ratio of $R^{*2}_{gR} /R^{*2}_{gL}$ with chain length under $\theta$-conditions.}
\label{fig:Rg2_theta}
\end{figure*} 

\section{Universal static properties}\label{sec:Uniratio_rings}

The universal scaling of equilibrium ratios involving the radius of gyration, $R_g$, and the mean stretch along the $x$-axis, $X$, for rings and the corresponding linear chains, as a function of solvent quality $z$ in the crossover regime, is investigated in this section. The radius of gyration is defined as
\begin{equation}\label{Eq:Rg2}
R_g^{2} \coloneqq \langle R_{g}^{2} \rangle = \frac{1}{2 N_b^2}
\sum_{p = 1}^{N_b} \sum_{q = 1}^{N_b} \langle  r_{p q}^{2}\rangle
\end{equation}
with ensemble averages being represented by the angular brackets, and $r_{pq} = \vert \bm{r}_q - \bm{r}_p \vert$ denoting the inter-bead distance. The size of a polymer chain is often also characterised by the mean-stretch of the chain along one of the coordinate axis. The stretch of a chain along any axis is defined as the maximum projected length of the chain along that axis. The mean-stretch of the chain along the $x$-axis is defined as $X = \left\langle \text{max}|x_{\mu}-x_{\nu}|\right\rangle$, where $x_{\mu}$ is the $x$-component of vector $\bm{r}_{\mu}$. This is an important property to study as it is regularly and easily measured in experiments involving stained DNA molecules~\cite{SmithChu,Hsiao2017,Sasmal2017,Tu2020} and is more readily accessible than $R_g$, particularly for large molecules.

In the $\theta$ solvent limit ($z=0$), the {non-dimensional} mean-squared radius of gyration $R_g^{*2}$ for both linear and ring polymers follows linear scaling with chain length, $N_b$, as shown in Fig.~\ref{fig:Rg2_theta}~(a), which corresponds to random walk statistics for both linear and ring polymers. Furthermore, it can be seen from Fig.~\ref{fig:Rg2_theta}~(b) that the mean-squared radius of gyration of a single-ring is half that of a linear chain under $\theta$ solvent conditions, which is in agreement with earlier theoretical predictions of the ratio $\left( R^\theta_{gR} /R^\theta_{gL} \right)^2 \left( =  {R^{\theta *}_{gR}}^2 /{R^{\theta *}_{gL}}^2  \right)$ for ideal Gaussian chains~\cite{Kramers1946,Zimm1949,Yamakawa1971}, and careful experimental observations carried out at the respective $\theta$-temperatures of the ring and linear polymer solutions~\cite{Roovers1985,Kovacs1987,Vlassopoulos2015}.

\begin{figure*}[tb]
  \begin{center}
		\begin{tabular}{cc}
		 \hspace{-1cm}
			\resizebox{8.9cm}{!}{\includegraphics*[width=4cm]{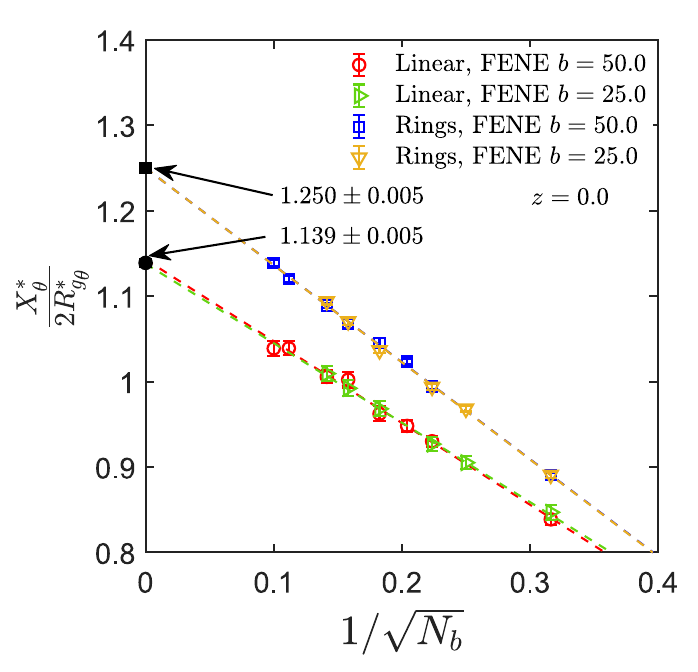}}   &
		 \hspace{-0.5cm} 	
			\resizebox{9.1cm}{!} {\includegraphics*[width=4cm]{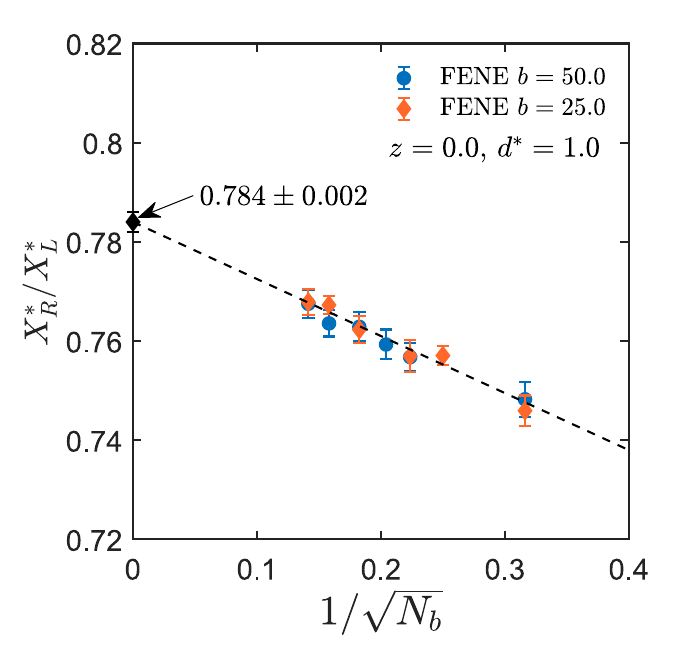}} \\
			(a) & (b) \\
		\end{tabular}
	\end{center}
	\vskip-15pt
\caption{(a) The ratio $X^*/2R^*_g$ as a function of $1/\sqrt{N_b}$ for linear and ring chains in $\theta$-solvents. (b) The ratio of the mean stretch of a ring to that of a linear chain, $X^*_R/X^*_L$, as a function of $1/\sqrt{N_b}$ in $\theta$-solvents. The extrapolated values in the limit of $N_b \rightarrow \infty$ are shown in the figure. The symbols are simulation data and the broken lines are linear fits to the data.}
\label{fig:Xratio_theta}
\end{figure*}  

\begin{figure*}[tbh]
  \begin{center}
		\begin{tabular}{cc}
		 \hspace{-1cm}
			\resizebox{9.1cm}{!}{\includegraphics*[width=4cm]{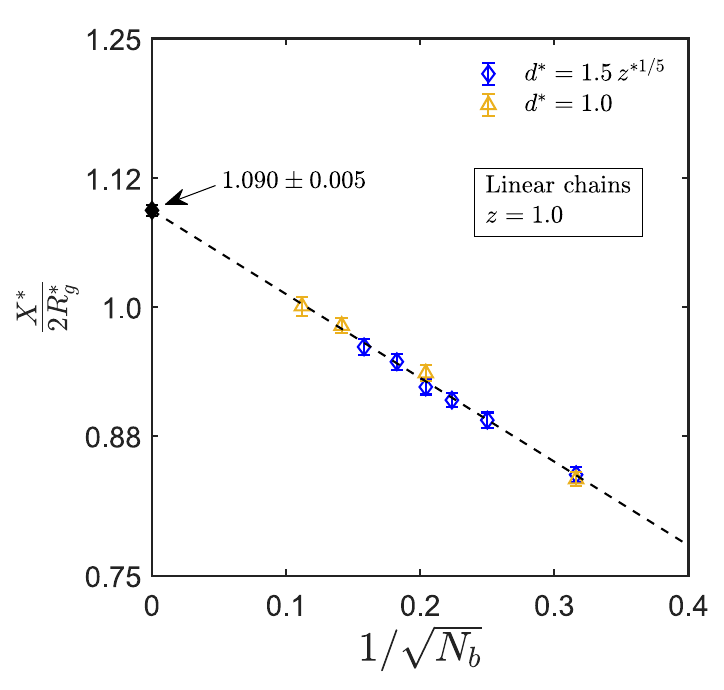}}   &
		 \hspace{-0.5cm} 	
			\resizebox{9cm}{!} {\includegraphics*[width=4cm]{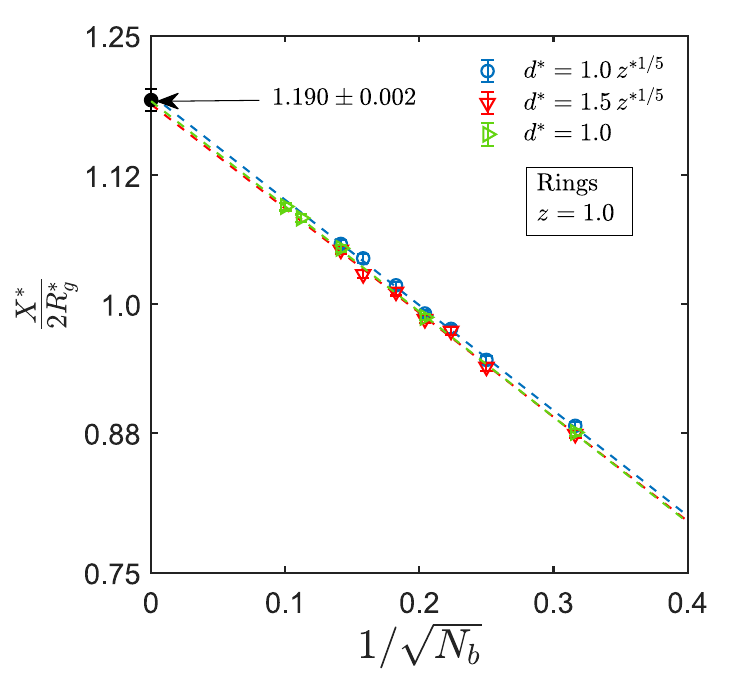}} \\
			(a) & (b) \\
		\end{tabular}
	\end{center}
	\vskip-15pt
\caption{Ratio $X^*/2R^*_g$ as a function of $N_b$ for (a) linear and (b) ring polymer solutions at solvent quality parameter $z=1.0$. Two different approaches are considered to specify the interaction range, $d^*$. The parameter $d^*$ is defined as $d^* = k\,z^{*1/5}$ in one approach, with $k=1.0$ and $1.5$, and a constant value $d^*=1.0$, is used in the other. The extrapolated values in the limit of $N_b \rightarrow\infty$ are $1.090\pm0.005$ and $1.190\pm0.002$ for linear and ring chains, respectively.}
\label{fig:XbyRgRP_z1}
\end{figure*}

\begin{figure}[ptbh]
  \begin{center}
		 \hspace{-1cm}
			\resizebox{9.5cm}{!}{\includegraphics*[width=4cm]{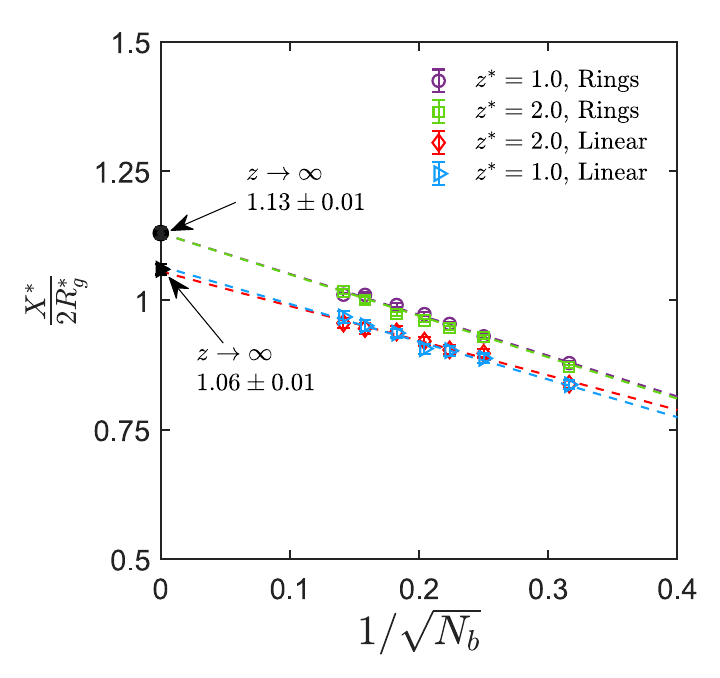}}  
	\end{center}
	\vskip-15pt
\caption{Ratio $X^*/2R^*_g$ as a function of $N_b$ for linear and ring polymers at two different values of $z^*$. The extrapolated values of the ratio in the limit of $z\rightarrow\infty$ for linear and ring chains are $1.06\pm 0.01$ and $1.13\pm 0.01$, respectively.}
\label{fig:XbyRgzinf}
\end{figure}

To investigate the effect of solvent quality on the non-dimensional mean stretch along the $x$-axis, $X^*$, and the non-dimensional radius of gyration, $R_g^*$, we have systematically increased the solvent quality parameter, $z$, and calculated the ratio $X/2R_g = X^*/2R_g^*$ for both rings and linear chains. Additionally, the ratio of mean stretch, $X_R/X_L = X_R^*/X_L^*$, of ring to linear (where, $X_R^*$ and $X_L^*$ are the respective non-dimensional mean stretch of a ring and linear chain along $x$-axis) is also investigated as a function of $z$. As discussed in greater detail below, since the solvent quality $z$ is dependent on chain length, the value of the parameter $z^*$ ($=z/\sqrt{N_b}$) for different chain lengths $N_b$ is adjusted in the successive fine-graining process such that $z$ is kept constant at fixed values of solvent quality in the crossover regime. However, in case of $\theta$-solvents, since $z=0$, simulations for different chain lengths are carried out with $z^*=0$. The universal values of the ratio $X^*/2R_g^*$ and $X_R^*/X_L^*$ are evaluated in the limit of infinite chain length by progressively increasing $N_b$ and extrapolating the values of these quantities to $N_b \rightarrow \infty$, as demonstrated in Figs.~\ref{fig:Xratio_theta} for systems under $\theta$-solvent conditions ($z=0$). Data is plotted as a function of $1/\sqrt{N_b}$ since it can be shown that leading order corrections to the infinite chain length limit, for properties that depend on hydrodynamic interactions~\cite{ottrab89} and on excluded volume interactions~\cite{prakash2001influence}, scale as $1/\sqrt{N_b}$. In Figs.~\ref{fig:Xratio_theta}~(a) and (b), the universality of $X^*/2R_g^*$ and $X_R^*/X_L^*$ under $\theta$-solvent conditions is established by computing the ratios for two different values of the FENE $b$-parameter, $b = 50.0$ and $b=25.0$. It is interesting to note that the values of both the ratios, for both linear and ring chains, are independent of the FENE $b$-parameter not only in the limit of infinite chain length but also for finite chain sizes. According to Fig.~\ref{fig:Xratio_theta}~(a) the universal values of the ratio $X^*/2R_g^*$ in the $\theta$-solvent limit for linear and ring polymers are found to be $1.139\pm0.005$ and $1.250\pm0.005$, respectively. These values are in close agreement with previously reported values of $1.132\pm0.005$ for linear chains obtained from BD simulations~\cite{Sunthar2005} and $1.1284$ and $1.2533$ for linear chains and rings, respectively, derived from analytical theory based on the Gaussian chain model~\cite{Zhu2016}. The universal value of the ratio $X_R^*/X_L^*$ under $\theta$-solvent conditions is found to be $0.784\pm0.002$, which is also in good agreement with the predicted value of $0.7854$ obtained from the Gaussian chain model~\cite{Zhu2016}. 

To calculate the universal values of the ratio of these static properties ($X^*/2R_g^*$ and $X_R^*/X_L^*$) in the crossover regime, $0<z<\infty$, a similar method of extrapolation is adopted. However, contrary to the trivial case of a $\theta$-solvent, the value of $z^*$ ($=z/\sqrt{N_b}$) is adjusted depending on chain length $N_b$, such that $z$ remains constant for a given solvent quality. An example of this method is shown in Figs.~\ref{fig:XbyRgRP_z1} for linear and ring polymer solutions at $z=1.0$. Here, universality of the ratio is established with respect to different values of the range of the EV interaction potential, $d^*$. As displayed in Figs.~\ref{fig:XbyRgRP_z1}, two different approaches are considered, where in one case $d^*$ is kept constant at a value of $1.0$, whereas, in the other $d^*$ is varied according to the relation $d^* = k\,z^{*1/5}$, where $k$ is a non-dimensional constant. In the latter case, it should be noted that both $z^*$ and $d^*$ change with $N_b$ in the fine-graining process. As discussed by \citet{KumarRavi}, this form of the relation chosen for $d^*$ is found to be computationally efficient in the context of BD simulations. Here, two different values of the constant $k$ have been used for linear and ring polymers, $k=1.0$ and $k=1.5$, as indicated in Figs.~\ref{fig:XbyRgRP_z1}. Interestingly, for all the different approaches considered here there exists a unique universal value of the ratio $X^*/2R_g^*$, which is $1.190\pm0.002$ for rings and $1.090\pm0.005$ for linear chains, at $z=1.0$. Furthermore, the values of the ratio are independent of $d^*$ for both finite and infinite chain lengths. Typically, for \textit{finite} chain sizes, the ratios of different static and dynamic properties or swelling ratios, are found to be dependent on the parameter $k$ or $d^*$~\cite{KumarRavi,Sunthar2006}, however, this is a rather interesting finding here which suggests that for the ring topology, the ratio $X^*/2R_g^*$ is independent of the parameter $k$ or $d^*$ for all values of $N_b$.  

\begin{figure*}[tbh]
  \begin{center}
		\begin{tabular}{cc}
		 \hspace{-1cm}
			\resizebox{9cm}{!}{\includegraphics*[width=4cm]{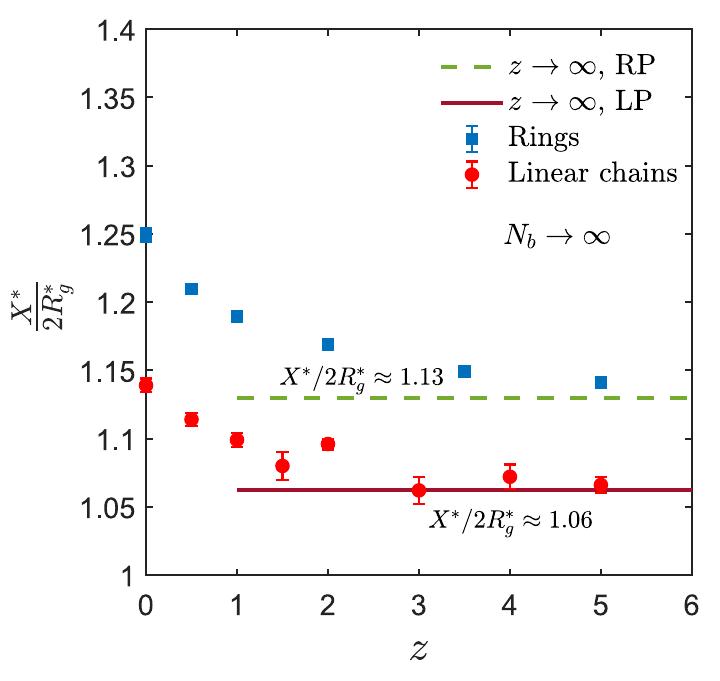}}   &
		 \hspace{-0.5cm} 	
			\resizebox{9cm}{!} {\includegraphics*[width=4cm]{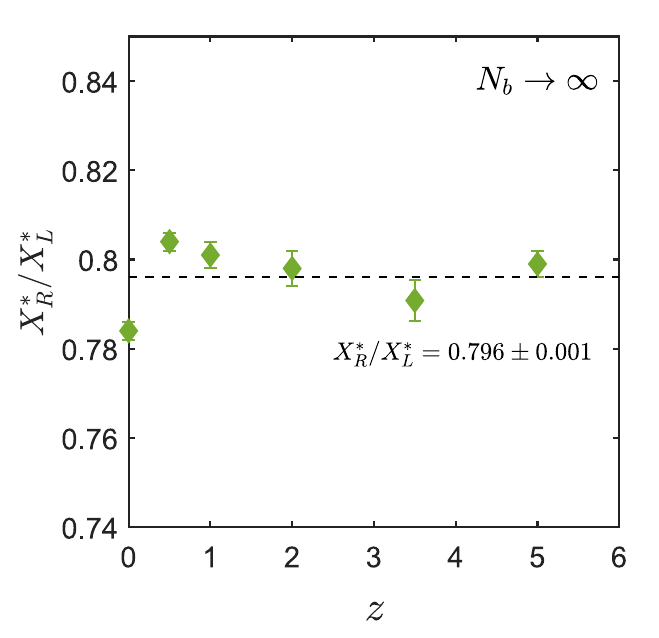}} \\
			(a) & (b) \\
		\end{tabular}
	\end{center}
	\vskip-15pt
\caption{(a) The ratio $X^*/2R^*_g$ in the crossover regime for linear and ring chains in the long chain limit. The solid and the dashed lines represents the value of $X^*/2R^*_g$ in limit of $z \rightarrow \infty$ for linear and ring polymers, respectively, derived from finite $N_b$ simulations data  at constant $z^*$, extrapolated to $N_b \to \infty$. (b) The ratio of the mean stretch of a ring and linear chain, $X^*_R/X^*_L$, in the crossover regime in the long chain limit. The dashed line indicates the average value of the ratio for large $z$.}
\label{fig:Xratio_v_z}
\end{figure*}

It is worth noting that since $z = z^*\sqrt{N_b}$, the asymptotic limit of $z\rightarrow \infty$ can be reached by carrying out simulations at a fixed value of $z^*$ for various chain lengths and extrapolating the measured property to $N_b\rightarrow\infty$. An example of this method is demonstrated in Fig.~\ref{fig:XbyRgzinf}, where the ratio $X^*/2R_g^*$ is computed for rings and linear chains at different values of $z^*$ and extrapolated to the limit of infinite chain length. We find that for both the chain architectures, the ratio $X^*/2R_g^*$ is independent of $z^*$ for finite chains and in the infinite chain length limit. The values of the ratio $X^*/2R_g^*$ for linear and ring chains in the asymptotic limit of $z\rightarrow \infty$ are $1.06\pm 0.01$ and $1.13\pm 0.01$, respectively, which are in good agreement with the values $1.045\pm0.005$ and $1.14\pm0.005$, obtained by \citet{Zhu2016} for linear chains and rings, respectively, by carrying out numerical simulations of the Kremer-Grest bead-spring model with a Weeks-Chandler-Andersen potential to represent excluded volume interactions in the athermal solvent limit. 

In Fig.~\ref{fig:Xratio_v_z}~(a), we show that for both ring and linear chains the ratio $X^*/2R_g^*$ is found to decrease with increasing solvent quality in the crossover regime and asymptotically tends to a constant value as $z \rightarrow \infty$. Note that the value of the ratio in the limit of $z \rightarrow \infty$ can be approached in two different ways; i.e., either by increasing $z$ asymptotically to infinity and estimating the value of the ratio as $z \rightarrow \infty$ (as shown in Fig.~\ref{fig:Xratio_v_z}~(a)), or by keeping $z^*$ constant and extrapolating to $N_b \rightarrow \infty$ (as displayed in Fig.~\ref{fig:XbyRgzinf}). While, typically, these two different approaches may not lead to the same result~\cite{KumarRavi}, here, we have observed that both these methods gives the same value for the ratio $X^*/2R_g^*$ in the limit of $z \rightarrow \infty$. It may also be noted that $X^*/2R_g^*$ for linear chains approaches the asymptotic limit faster (i.e., at smaller values of $z$) than for rings, and the value of the ratio is larger for rings than for linear chains at any value of $z$. 

The ratio $X_R^*/X_L^*$, on the other hand, rapidly approaches a constant value of about $0.796\pm0.001$ by fairly small values of the solvent quality $z$, as displayed in Fig.~\ref{fig:Xratio_v_z}~(b). This asymptotic value is close to that of $0.8$, reported by \citet{Zhu2016} using numerical simulations of the Kremer-Grest bead-spring model in good solvents. It should be noted that \citet{Zhu2016} have only reported values of $X_R^*/X_L^*$ in the two limiting cases, i.e., $X_R^*/X_L^* = 0.7854$ for $\theta$ solvents obtained using analytical theory, and $X_R^*/X_L^* = 0.8$ for athermal solvents obtained using numerical simulations. Here, the value of the ratio has been determined throughout the crossover regime. Since $X_R^*/X_L^* < 1$ for all values of $z$, a ring chain is always less stretched than a linear chain having the same number of monomers or molecular weight at identical values of $z$. Within the resolution of the error bars, Fig.~\ref{fig:Xratio_v_z}~(b) suggests that rather than increasing monotonically from the $\theta$ to the athermal solvent limit value, the ratio $X_R^*/X_L^*$ first increases above the asymptotic value at small values of $z$ before decreasing to the asymptotic value. The difference is, however, in the second decimal place, and for all practical purposes the value of $X_R^*/X_L^*$ maybe considered nearly independent of solvent quality.

\section{Universal dynamic properties and swelling ratios}\label{sec:UniratioHI_rings}

Contrary to the mean-squared radius of gyration or mean stretch, which are equilibrium static properties, the hydrodynamic radius, $R_H$, is a dynamic property since it is related to the diffusivity of polymer chains~\cite{RubColby2003,Sunthar2006}. In this section we discuss the methodology to compute the hydrodynamic radius of rings and calculate different universal ratios involving $R_H$. The hydrodynamic radius and the mean-stretch (discussed in Section.~\ref{sec:Uniratio_rings}) are two properties of polymer solutions (particularly DNA solutions) that are more conveniently and regularly measured in experiments than the radius of gyration, $R_g$. This makes it worthwhile to compute the ratio involving the mean stretch, $X$, and the hydrodynamic radius, $R_H$, as a function of solvent quality for rings and compare them with that of linear chains. For completeness we have also studied the ratio of $R_H$ to $R_g$ as a function of solvent quality. The swelling behaviour of the gyration radius, hydrodynamic radius and mean stretch relative to their values in the $\theta$-state i.e.,  $\alpha_g = R_g/R_g^{\theta}$, $\alpha_H=R_H/R_H^{\theta}$, and $\alpha_X=X/X_{\theta}$, has also been investigated in this section.

\subsection{Calculation of the long-time diffusivity and the hydrodynamic radius}\label{subsec:diff_Rh}

The hydrodynamic radius is inversely proportional to the long-time diffusivity $D$ of a polymer chain as defined by the following equation
\begin{equation}
D = \frac{k_B T}{6\pi\eta_s}\frac{1}{R_H} 
\end{equation}
where the quantities $k_B$, $T$ and $\eta_s$ have been defined earlier below Eq.~(\ref{GovEqn}). The long-time diffusivity, $D$, is typically computationally determined from the mean-squared displacement of the center of mass of a polymer chain, given by
\begin{equation}
\left\la \left( \Delta r_{\text{cm}} \right)^2 \right\ra = 6\,Dt 
\end{equation}
where $\Delta r_{\text{cm}} (t) $ is the displacement of the centre of mass as a function of time $t$. However, the calculation of $D$ from the mean-squared displacement requires the simulation of long trajectories and tends to be error prone~\cite{Liu2003}. Since the prediction of universal properties requires the extrapolation of finite chain data, it is vitally important to obtain finite chain results with relatively small error bars at a reasonable computational cost. An alternative method to estimate the long-time diffusivity $D$ involves using Fixman's formula,  which is based on the Kirkwood expression for the short-time diffusivity. Fixman's formula is given by $D = D_K - D_1$, where $D_K$ is the short-time diffusivity calculated from the Kirkwood expression and $D_1$ represents intra-molecular dynamic correlations. The expression for $D_K$ is,
\begin{equation}
D_K = \frac{D_0}{N_b} + \frac{k_B T}{6\pi\eta}\frac{1}{R_I}
\end{equation}
where $D_0 = k_BT/\zeta$ is the diffusivity of a single bead and $R_I$ (which is a static property) is the inverse radius defined by,
\begin{equation}
\frac{1}{R_I} = \frac{1}{N_b^2}\,\sum\limits_{\substack{\mu,\nu=1 \\ \mu\ne\nu}}^{N_b} \left \la \frac{1}{r_{\mu\nu}} \right \ra
\end{equation}
By appropriate non-dimensionalisation, the dimensionless dynamic correlation function $D^*_1$ is defined in terms of the auto correlation function of the quantity $A^*_i$, 
\begin{equation}\label{Eq:FixmanCorr}
D^*_1 = \frac{1}{3N_b^2} \sum\limits_{i=1}^3 \int\limits_{0}^{\infty}dt\,\la A^*_i(0)\,A^*_i(t)\ra
\end{equation}
where $A^*_i = \displaystyle\frac{1}{4}\sum\limits_{\mu=1}^{N_b}\sum\limits_{\nu=1}^{N_b}\sum\limits_{j=1}^3D^*_{\mu\nu ij}F^*_{\nu j}$. The quantity $D^*_{\mu\nu ij}$ is the $ij^{th}$ component of the diffusion tensor for bead pair $(\mu,\nu)$ and $F^*_{\nu j}$ defines the force acting on bead $\nu$ in the $j$-direction. After non-dimensionalisation, the following equation relating the dimensionless diffusivity and the dimensionless hydrodynamic radius can be derived,
\begin{equation}
D^* = \frac{h^*\sqrt{\pi}}{4 R_H^*}
\end{equation}
where $D^* = D (\lambda_H/l_H^2)$.

A careful discussion of the implications of calculating $D^*$ via the displacement of the centre of mass versus Fixman's formula is given in \citet{Liu2003}. In particular, it should be noted that computer simulations often report the value of $D^*_K$ as the diffusivity rather than $D^*$, since the difference in magnitude is small for finite chains~\cite{Liu2003,Sunthar2006}. On the other hand, Fixman's formula makes it clear that $D^*_K$ is a static property since it is only dependent on $R^*_I$, while $D^*$ is a dynamic property because of the correction term accounting for intra-molecular correlations, $D^*_1$. As will be discussed in greater detail in Section~\ref{subsec:UniSwell}, this has very significant implications for the prediction of the swelling of the hydrodynamic radius, $\alpha_H$ in the long chain limit.

\begin{figure}[ptbh]
  \begin{center}
		 \hspace{-1cm}
			\resizebox{9cm}{!}{\includegraphics*[width=4cm]{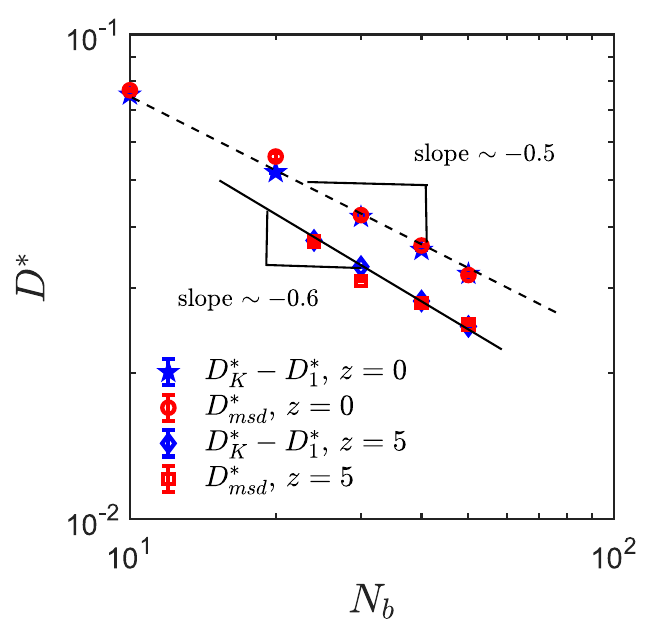}}  
	\end{center}
	\vskip-15pt
\caption{Scaling of the non-dimensional long-time diffusivity $D^*$ with chain length $N_b$, at $z=0$ ($\theta$ solvent) with $h^*=0.2$, and at $z=5$ (good solvent) at $h^*=0.24$, for rings. Here, $D^*$ is computed by the two methods discussed in the text. The dashed and solid lines indicate the scaling exponents for the two solvent qualities.}
\label{fig:DvsDmsd}
\end{figure} 

Previously, it has been shown that the non-dimensional long-time diffusivities computed for linear chains from the mean-squared displacement ($D^*_{msd}$) and Fixman's formula ($D^*_K- D^*_1$) are consistent with each other~\cite{Liu2003,Sunthar2006}. Fig.~\ref{fig:DvsDmsd} confirms that the values of the long-time diffusivity for rings calculated by both these methods are also consistent with each other, in both the $\theta$ and good solvent limits. An illustration of the procedure for calculating the non-dimensional diffusivity using both these methods is provided in the Appendix. Values of $D^*_K$, $D^*_1$ and $D^*$ estimated from simulations for various chain lengths and solvent qualities have been tabulated in the Appendix. As can be seen from Table~\ref{tab:D1Dk_data}, while the ratio $D^*_1/D^*_K$ is less than 3.5\% for the chain lengths that have been examined here, its value is increasing with increasing chain length. Fig.~\ref{fig:DvsDmsd} also shows that, similar to linear polymers, the non-dimensional diffusivity of rings scales with chain length, $N_b$, as $N_b^{-0.5}$ in $\theta$ solvents and $N_b^{-0.6}$ in athermal solvents. These results are in agreement with the diffusivity scaling observed in the experiments on circular DNA by \citet{Robertson2006}.

\subsection{Universal ratios involving the hydrodynamic radius}\label{subsec:UniRatioRh}

\begin{figure}[pt]
  \begin{center}
		 \hspace{-1cm}
			\resizebox{9cm}{!}{\includegraphics*[width=4cm]{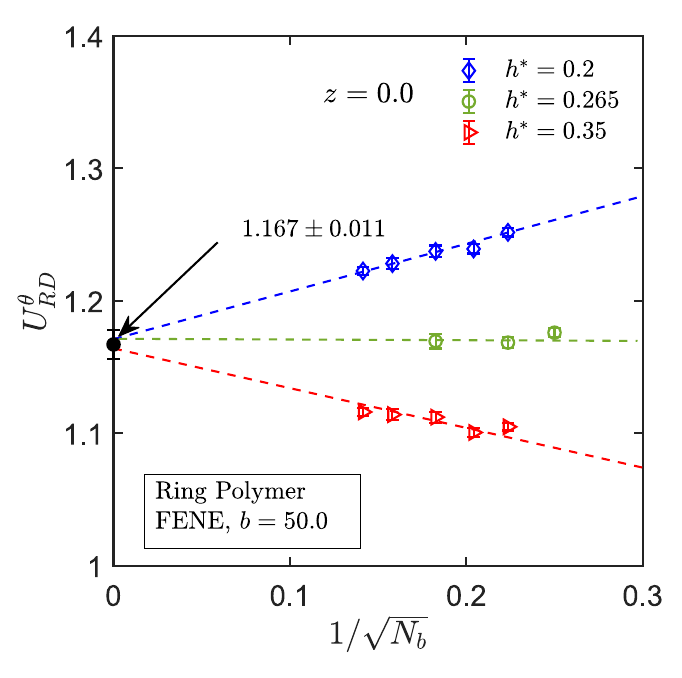}}  
	\end{center}
	\vskip-15pt
\caption{The ratio $U_{\text{RD}}$ as a function of $1/\sqrt{N_b}$ for rings in $\theta$-solvents, $z=0$, at several values of $h^*$. The extrapolated value of the ratio in the limit of $N_b \rightarrow \infty$ is universal and equal to {$1.167\pm0.011$}. The symbols are simulation data and the broken lines are linear fit to the data.}
\label{fig:URD_z0_RP}
\end{figure}

With this necessary background to compute the non-dimensional hydrodynamic radius, $R^*_H$, from the non-dimensional long-time diffusivity, we examine the variation of the universal ratio, $U_{\text{RD}} = R_g/R_H = R^*_g/R^*_H$, as a function of the solvent quality $z$ in this subsection. It is well established~\cite{ottrab89,Kroger2000,Sunthar2006} that for fixed values of the solvent quality parameter, $z$, the ratio $U_{\text{RD}}$ is independent of the strength of hydrodynamic interactions ($h^*$) in the non-draining limit, $h\rightarrow\infty$, where $h=h^*\sqrt{N_b}$. While \citet{Sunthar2006} have shown that this universality holds true for linear chains, here we establish the validity of this universal behaviour for rings. Normally, in simulations with the FENE spring force law, the quantity ${\tilde h}^*$ (which is a function of the FENE $b$ parameter) is kept constant instead of $h^*$. This is important in simulations where finite chain data is extrapolated to the number of Kuhn steps in the polymer chain (rather than to infinity), with the contour length (corresponding to a specific number of Kuhn segments) being kept unchanged with an increase in the number of beads $N_b$ in the {successive fine-graining} process~\cite{Sunthar2005,Sasmal2017}. However, since the present study considers an infinite number of Kuhn segments (corresponding to infinite chain length) and a constant value of $b$, it is sufficient to keep $h^*$ constant for simulations at different values of $N_b$. In Fig.~\ref{fig:URD_z0_RP} we calculate $U_{\text{RD}}$ as a function of chain length for several different values of $h^*$ at $z = 0$. In the extrapolated limit of $N_b \rightarrow \infty$, curves for different values of $h^*$ converge to a common value, $1.167\pm 0.011$. This is the universal value of the ratio $U_{\text{RD}}$ for rings, which is independent of the parameter $h^*$, at $z=0$. For linear polymers the value of this ratio is found to be $1.375\pm0.004$ (not shown here), similar to previous predictions by Sunthar and Prakash~\cite{Sunthar2005} and \citet{Kroger2000}. \citet{Uehara2016} have estimated the value of  $U_{\text{RD}}$ to be $1.253\pm 0.013$ for rings, which is significantly higher than the value predicted here. However, it should be noted that they have computed $R^*_H$ from the Kirkwood expression without Fixman's correction, and such a treatment is valid only for the estimation of short-time diffusivities.

The difference in the values of the ratio $U_{\text{RD}}$ for different values of $h^*$ for finite chains is attributed to leading order corrections to the estimate of the infinite chain length limit of $U_{\text{RD}}$, and it has been shown to scale with chain length as follows~\cite{Kroger2000}
\begin{equation}\label{Eq:URD}
U_{\text{RD}}(h^*,N_b) = \tilde{U}^{\infty}_{\text{RD}} + \frac{c_{\text{RD}}}{\sqrt{N_b}}\left(\frac{1}{h^*_{\text{RD}}}-\frac{1}{h^*}\right) + \mathcal{O}\left(\frac{1}{N_b}\right)
\end{equation}
where $\tilde{U}^{\infty}_{\text{RD}}$ is the value of the ratio in the limit of infinite chain length, and $c_{\text{RD}}$ and $h^*_{\text{RD}}$ are constants. From Eq.~(\ref{Eq:URD}) it is clear that one can find the fixed point value of $h^*$ at which the leading order correction term drops out and the ratio converges more quickly to its universal value, by solving the equation for $c_{\text{RD}}$ and $h^*_{\text{RD}}$ with the help of simulation data. Following this method \citet{Kroger2000} found the fixed point value of the hydrodynamic interaction parameter to be $h^* = h^*_{\text{RD}}=0.267$ for linear polymers. For rings we have found that the fixed point occurs at $h^* = h^*_{\text{RD}} \approx 0.265$. This is illustrated in Fig.~\ref{fig:URD_z0_RP} by showing the ratio $U_{\text{RD}}$ is nearly independent of chain length $N_b$ at $h^*=0.265$. The ratio $U_{\text{RD}}$ is computed here for a range of values of $z$ using $h^*=0.24$, which is close to the fixed point value.

\begin{figure}[pb]
 \begin{center}
		 \hspace{-1cm}
			\resizebox{9cm}{!}{\includegraphics*[width=4cm]{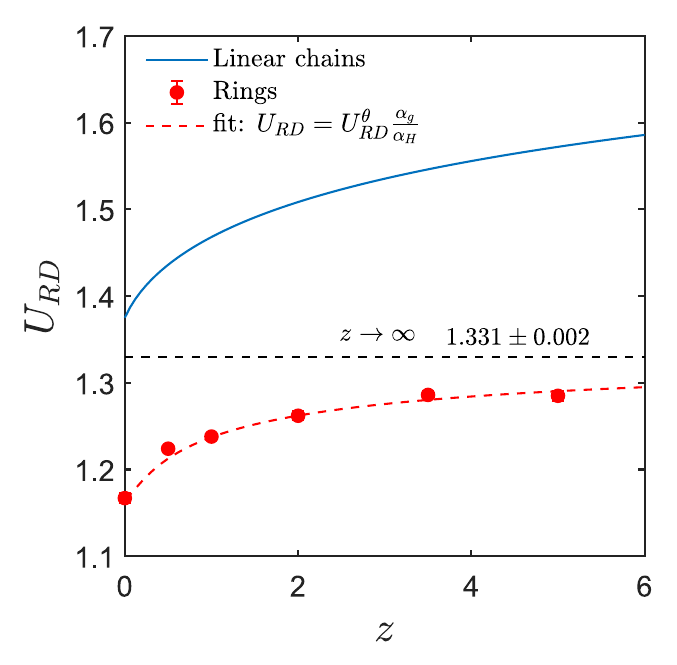}}  
	\end{center}
	\vskip-15pt
\caption{The ratio $U_{\text{RD}}$ as a function of $z$ for rings and linear chains in the long chain limit. The red dashed curve corresponds to the equation fitting $U_{\text{RD}}$ for rings. The horizontal dashed line represents the value of the ratio in the limit of $z \rightarrow \infty$ for ring polymers, which is obtained by keeping $z^*$ constant and extrapolating finite chain data to $N_b\rightarrow \infty$.}
\label{fig:URD_vs_z}
\end{figure}

\begin{table*}[ptbh]
\caption{Coefficients and constants for the expression of the swelling ratio given in Eq.~(\ref{Eq:Swellratio}) for linear chains and rings. Quantities without error bars are those that have been fixed at the indicated values, and act as constraints to the fit for the remaining parameters.}
\label{Tab:SRringlinear}
\centering
{\footnotesize
\begin{tabular}{c|c|c|c|c|c|c}
\hline
 & \multicolumn{3}{c}{Linear}  & \multicolumn{3}{|c}{Rings}  \\
 \hline
 \hline
 const &$\alpha_g$&$\alpha_H$&$\alpha_X$      &$\alpha_g$&$\alpha_H$&$\alpha_X$\\
 \hline
 \hline
 \multirow{1}{*}{$a$}& $9.5286$ & 9.528 & 9.528  & $10.91\pm 0.65$ & $10.9$ & $10.9$ \\
 \hline
\multirow{1}{*}{$b$}&$19.48\pm 1.28$ & 19.48& 19.48  & $33.35\pm 0.62$ & $11.79\pm 1.06$ & $11.79\pm 1.06$   \\
   \hline
 \multirow{1}{*}{$c$}& $14.92\pm 0.93$ & 14.92& 14.92  & $11.61\pm 0.71$ & $1.967\pm 0.315$ & $1.967\pm 0.315$ \\
 \hline
 \multirow{1}{*}{$m$}& $0.1339  \pm 0.0006$ & $0.0995 \pm 0.0014$ & $0.114 \pm 0.0013$  & $0.1469\pm 0.0007$ & $0.1469\pm 0.0013$  & $0.1469\pm 0.0013$  \\
 \hline        
\end{tabular}
}
\end{table*}

In Fig.~\ref{fig:URD_vs_z} we show the variation of $U_{\text{RD}}$ as a function of $z$ for rings and linear chains. Note that the universal ratio, $U_{\text{RD}}$, for linear chains is computed from the formula $U_{\text{RD}} = \alpha_g U_{\text{RD}}^{\theta}\alpha_H^{-1}$, where $\alpha_g = R_g/R_g^{\theta} = R_g^*/R_g^{*\theta}$ and $\alpha_H=R_H/R_H^{\theta} = R_H^*/R_H^{*\theta}$ are the swelling ratios of the gyration radius and the hydrodynamic radius, respectively. These swelling ratios are given by the following functional form~\cite{Schafer,KumarRavi},
\begin{equation}
f(z) = (1 + az + bz^2 + cz^3)^{m/2}
\label{Eq:Swellratio}
\end{equation}
where $f(z)$ represents the specific swelling ratio ($\alpha_g$ or $\alpha_H$) and $a$, $b$, $c$ and $m$ are fitting parameters, the values of which are given in Table~\ref{Tab:SRringlinear} (the entries for rings and $\alpha_X$ for linear chains  are discussed subsequently). The expression in Eq.~(\ref{Eq:Swellratio}) has been shown to provide an excellent fit to experimental data for both $\alpha_g$ and $\alpha_H$ for a wide variety of polymer solvent systems, and to data from BD simulations~\cite{Fujita,Tominaga2002,KumarRavi,Sunthar2006,Sharad2014}. 

\begin{figure*}[pbth]
  \begin{center}
		\begin{tabular}{cc}
		 \hspace{-1cm}
			\resizebox{9cm}{!}{\includegraphics*[width=4cm]{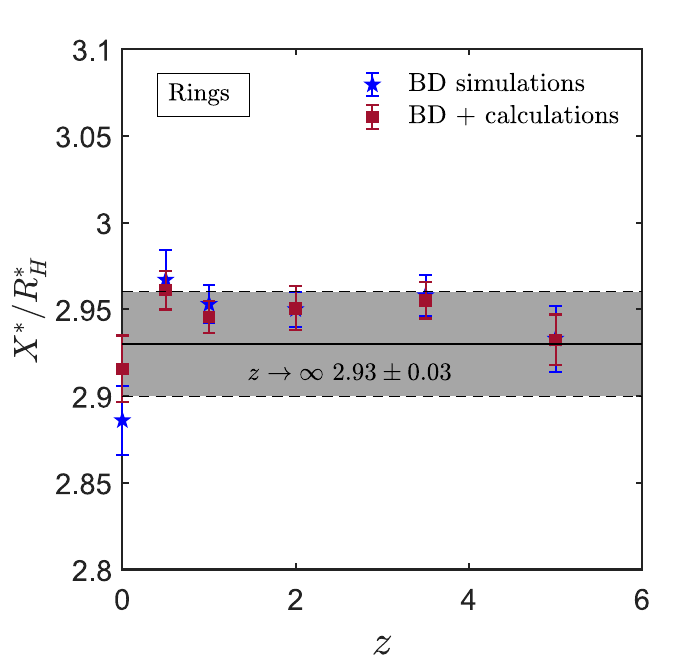}}   &
		 \hspace{-0.5cm} 	
			\resizebox{9cm}{!} {\includegraphics*[width=4cm]{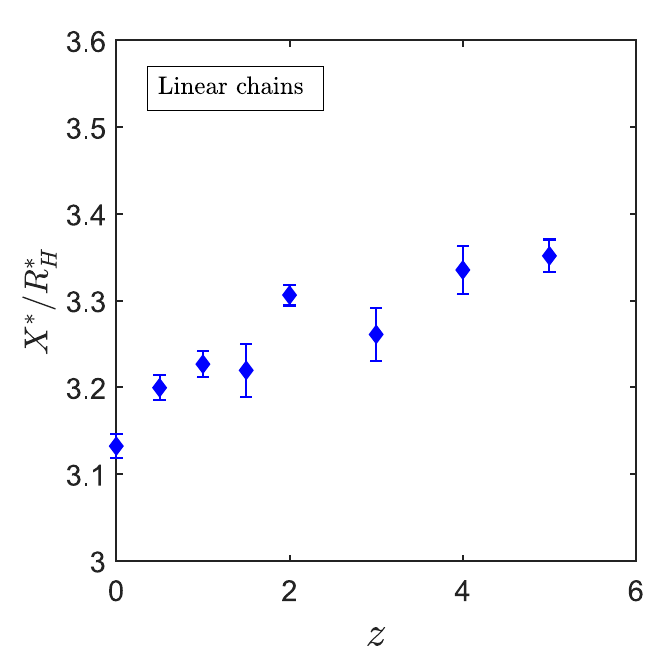}} \\
			(a) & (b) \\
		\end{tabular}
	\end{center}
	\vskip-15pt
\caption{The ratio $X/R_H = X^*/R_H^*$ as a function of $z$ for (a) rings and (b) linear chains in the long chain limit. The solid line (bounded by dashed lines) represents the value of the ratio in limit of $z \rightarrow \infty$ for ring polymers, where the $z \rightarrow \infty$ limit is obtained by keeping $z^*$ and extrapolating finite chain data to the limit of $N_b\rightarrow \infty$.}
\label{fig:XbyRh_vs_z}
\end{figure*}

It is apparent from Fig.~\ref{fig:URD_vs_z} that in the limit $z\rightarrow\infty$, the ratio $U_{\text{RD}}$ for rings in the crossover regime appears to be converging asymptotically to the value obtained by keeping $z^*$ constant and extrapolating finite chain data to $N_b\rightarrow \infty$. Moreover, the ratio $U_{\text{RD}}$ appears to grow more rapidly with $z$ for linear chains when compared to rings. It should be noted that in the case of linear chains, since the value of exponent $m$ for $\alpha_g$ is quite a bit greater than that for $\alpha_H$ (see  Table~\ref{Tab:SRringlinear}), the ratio $\alpha_g/\alpha_H$ never reaches an asymptotic value in the limit of $z\rightarrow\infty$. This is an important difference in the behaviour of rings and linear polymers. While $\alpha_g$ is a static property, $\alpha_H$ is a dynamic property, and for linear chains, the difference in behaviour of these properties is consistent with the values of the static and dynamic scaling exponents $\nu$ observed for the radius of gyration ($R_g \sim M^{0.59}$) and the hydrodynamic radius ($R_H  \sim M^{0.57}$), respectively, in the limit $z\rightarrow \infty$~\cite{deGennes79,Sunthar2006}. The dashed red line in Fig.~\ref{fig:URD_vs_z} that passes through the simulation data for rings, is drawn using the expression $U_{\text{RD}} = \alpha_g U_{\text{RD}}^{\theta}\alpha_H^{-1}$, where $\alpha_g$ and $\alpha_H$ for rings  (discussed below in Section~\ref{subsec:UniSwell}) have been calculated using the function $f(z)$ in Eq.~(\ref{Eq:Swellratio}), with the fitting parameters $a$, $b$, $c$ and $m$ given in Table~\ref{Tab:SRringlinear}. Since the value of exponent $m$ is the same for $\alpha_g$ and $\alpha_H$ in the case of rings, the ratio $\alpha_g/\alpha_H$  reaches an asymptotic value in the limit of $z\rightarrow\infty$, validating the suggestion from simulation data that the ratio $U_{\text{RD}}$ for rings converges to a constant value in this limit.

Another quantity of interest is the ratio of mean stretch, $X$, to the hydrodynamic radius, $R_H$. Following a similar method of extrapolation to infinite chain length, the universal values of the ratio $X/R_H = X^*/R_H^*$ are computed for a range of values of solvent quality parameter, $z$. As seen from Fig.~\ref{fig:XbyRh_vs_z}~(b), $X^*/R_H^*$ for linear chains has an increasing trend with $z$, whereas Fig.~\ref{fig:XbyRh_vs_z}~(a) for rings suggests that within the resolution of the error bars, the ratio is practically independent of solvent quality. In the case of ring polymers two methods have been adopted to estimate the ratio $X^*/R_H^*$. One is based on direct calculation of the ratio from BD simulations at different chain lengths and then extrapolating to infinite chain length and the second is based on the formula $X^*/R_H^* = \left( X^*/R_g^*\right) U_{\text{RD}}$. According to Fig.~\ref{fig:XbyRh_vs_z}~(a), both the methods give similar results within numerical accuracy. The near independence of the ratio with solvent quality for polymeric rings is an important finding which suggests that for rings the swelling in mean stretch and hydrodynamic radius is identical. This fact is further demonstrated in the swelling curve discussed later. Contrary to the case for linear polymers, it appears that the absence of any terminal ends in the ring architecture leads to $X^*$ and $R_H^*$ swelling by equal proportions. Interestingly, as can be seen from Fig.~\ref{fig:Xratio_v_z}~(a), the ratio $X^*/2R_g^*$ is not constant, but depends on the solvent quality $z$ for both rings and linear polymer solutions.

\subsection{Universal swelling ratios and the \textit{g-factor}}\label{subsec:UniSwell}

\begin{figure}[ptbh]
  \begin{center}
		 \hspace{-1cm}
			\resizebox{9.5cm}{!}{\includegraphics*[width=4cm]{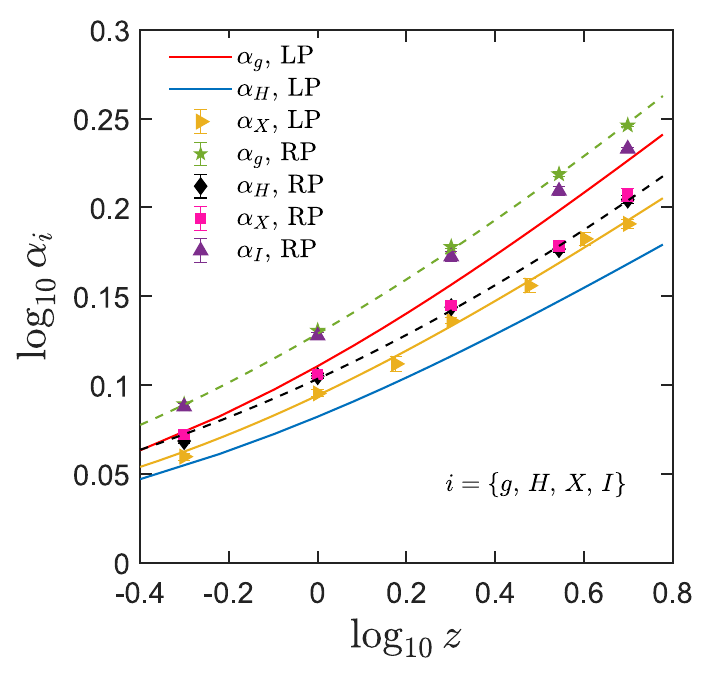}}  
	\end{center}
	\vskip-15pt
\caption{Swelling ratio $\alpha_g$, $\alpha_H$, $\alpha_X$ and $\alpha_I$ (displayed only for rings) plotted as a function of solvent quality, $z$, for linear (LP) and ring (RP) polymers. A common $y$-axis label $\alpha_i$, where $i = \{g, H, X, I \}$ is used to represent all the swelling functions.}
\label{fig:SwellRatio}
\end{figure}

The swelling of the radius of gyration and the hydrodynamic radius relative their values in a $\theta$-solvent, i.e., $\alpha_g$ and $\alpha_H$, respectively, are known to follow universal curves for linear polymers~\cite{KumarRavi,Sunthar2006}. However, for rings these swelling ratios have not been carefully studied in the thermal crossover regime so far. Here, we investigate the swelling behaviour of polymeric rings and compare the results with their linear chain counterparts. Fig.~\ref{fig:SwellRatio} displays the swelling ratios $\alpha_g$, $\alpha_H$ and $\alpha_X$ (the swelling of the mean stretch) as a function of $z$ for rings and linear chains. Symbols represent data for the various swelling ratios obtained by the current simulations. The only ratio computed here for linear chains is $\alpha_X$. Simulations to compute $\alpha_g$ and $\alpha_H$ for linear chains were not carried out since it has been previously established that the functional form for $f(z)$ provides an excellent fit to simulation data. For these ratios, the solid curves in Fig.~\ref{fig:SwellRatio} are drawn by inserting the appropriate fitting parameters given in Table~\ref{Tab:SRringlinear} into the functional form in Eq.~(\ref{Eq:Swellratio}). The dashed lines through the symbols for the ring polymer swelling ratios (and the solid line through the symbols for $\alpha_X$ for linear chains) are also fits using Eq.~(\ref{Eq:Swellratio}), with the fitting parameters given in Table~\ref{Tab:SRringlinear}. For linear polymers there is a clear distinction among the three swelling ratios whereas for rings the swelling of mean stretch, $\alpha_X$, and hydrodynamic radius, $\alpha_H$, are identical, consistent with the observation in Fig.~\ref{fig:XbyRh_vs_z}~(a) of the near independence of the ratio $X^*/R_H^*$ from $z$ for rings. On the other hand, while the curves for  $\alpha_g$ and $\alpha_H$ for rings are distinct from each other, they appear to become parallel for large values of $z$, which is expected since they have the same value of the exponent $m$. In other words, the swelling curves for rings suggest that static and dynamic scaling are identical in the case of rings.

\citet{Sunthar2006} have previously shown that for linear chains, the swelling of the inverse radius $\alpha_I = R_I^*/R_I^{*\theta}$ as a function of $z$ is identical to that of the swelling of the radius of gyration $\alpha_g$, since they both represent the swelling of static properties. This result helped establish that the experimentally observed difference between $\alpha_g$ and $\alpha_H$ could only be captured by simulations when $R_H^*$ is used to calculate the hydrodynamic radius rather than $R_I^*$, or equivalently, the long-time value $D^*$ is used to calculate the diffusivity rather than the short-time value $D_K^*$. The small difference between $D^*$ and $D_K^*$ observed for finite linear chains becomes significant in the infinite chain length limit. We have examined the same question here by computing the swelling behaviour of the inverse radius $\alpha_I$ for rings. The difference between $D^*$ and $D_K^*$ (for the finite chain lengths examined here) is small for rings as well, as can be seen from the expression $D^*/D_K^* = 1 - D_1^*/D_K^*$ with $D_1^*/D_K^* < 3.5 \%$, as discussed previously, and displayed in Table~\ref{tab:D1Dk_data} in the Appendix. Fig.~\ref{fig:SwellRatio} shows that exactly as in the case for linear chains, the swelling of $\alpha_g$ and $\alpha_I$ is identical for rings, and the small difference between $D^*$ and $D_K^*$ at finite chain lengths becomes amplified in the long chain limit, leading to distinct curves for $\alpha_g$ and $\alpha_H$. In contrast to the case for linear chains, however, the static and dynamic exponents for the scaling of chain size with molecular weight are identical to each other for rings, in the athermal solvent limit. Clearly, these subtle points would be difficult to appreciate without the consideration of intra-molecular dynamic correlations highlighted through Fixman's formula.

The computation of the universal ratio $U^\theta_\text{RD}$ and the swelling curve $\alpha_H$ for rings provides a means for indirectly estimating the $\theta$-temperature $\Theta_R$ for any ring polymer-solvent system from a single measurement of the hydrodynamic radius, if the chemistry-dependent constant $k_0$ for the linear polymer-solvent system is known. This can be seen from the following argument. Firstly, the radius of gyration for rings at the $\theta$-temperature can be estimated from the expression $R^\theta_{gR} = R^\theta_{gL} /\sqrt{2} =  b_k \sqrt{N_k/(6 \cdot 2)}$. The values of the number of Kuhn steps $N_k$ and the Kuhn step length $b_k$ are commonly known for many linear polymers~\cite{RubColby2003} and for linear $\lambda$-phage DNA~\cite{Pan2014a}. The hydrodynamic radius for rings at  $\Theta_R$ is consequently, $R^\theta_{HR} = R^\theta_{gR} /{U^\theta_\text{RD}}_R = (b_k/1.167) \sqrt{N_k/12} $. Assuming that the experimentally measured value of the hydrodynamic radius of a ring polymer with molecular weight $M_R$, at some temperature $T_R^\dag$, is $R_{HR}^{\dag}$, one can calculate the swelling ratio $\alpha_R^\dag = R^\dag_{HR}/R^\theta_{HR}$. The value of solvent quality $z^\dag$ at which this swelling would occur can be read from the universal swelling curve in Fig.~\ref{fig:SwellRatio}, from which one can calculate $\Theta_R$ using the expression, $z^\dag = k_0\,(1- {\Theta_R}/T^\dag_R)\sqrt{M_R}$, since all the other quantities are known. For instance, this procedure could be used to estimate the $\theta$-temperature for circular $\lambda$-phage DNA, since the value of the chemistry-dependent constant $k_0$ for linear $\lambda$-phage DNA in the Tris-Ethylenediaminetetraacetic acid Buffer, which is a commonly used solvent in experiments involving DNA solutions, has been reported in Refs.~\cite{Sharad2014} and~\cite{Pan2014a}.

In addition to the individual swelling behaviour of rings and linear chains, we have also investigated the dependence of the ratio $R^{2}_{gR}/R^{2}_{gL} = R^{*2}_{gR}/R^{*2}_{gL}$ (or the \textit{g-factor}) and the ratio of hydrodynamic radii, $R_{HR}/R_{HL} = R^*_{HR}/R^*_{HL}$, on the solvent quality, $z$. The effect of solvent quality on the ratio of mean stretch of rings to linear chains has already been addressed in section~\ref{sec:Uniratio_rings}. 

\begin{figure}[ptbh]
  \begin{center}
		 \hspace{-1cm}
			\resizebox{9cm}{!}{\includegraphics*[width=4cm]{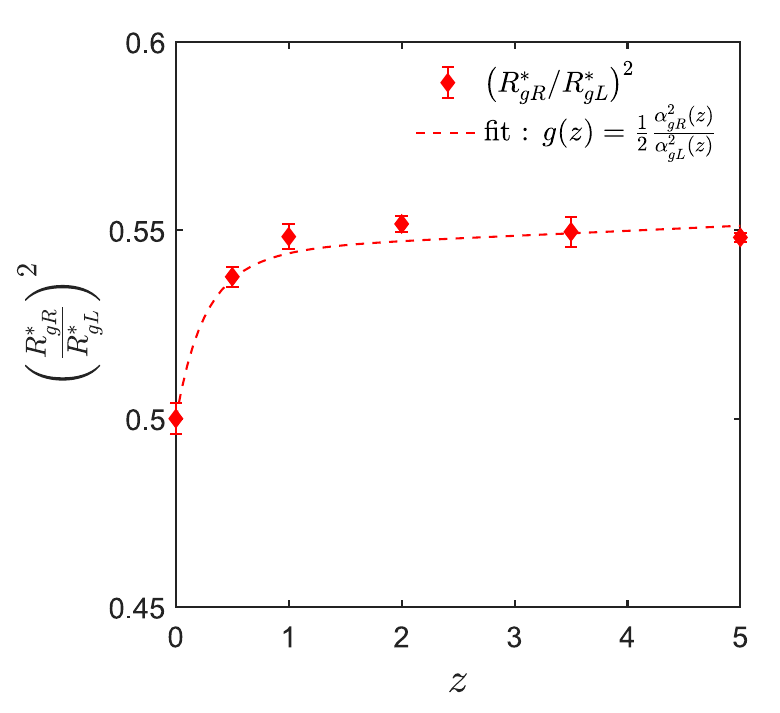}}  
	\end{center}
	\vskip-15pt
\caption{The  \textit{g-factor} $R^{2}_{gR}/R^{2}_{gL} = R^{*2}_{gR}/R^{*2}_{gL}$ plotted as a function of solvent quality, $z$. The symbols are from simulations and the dashed-line is given by the expression $g(z)=0.5 \left({\alpha_{gR}^2(z)}/{\alpha_{gL}^2(z)}\right)$, where the fitting parameters in Table~\ref{Tab:SRringlinear} are used for the swelling ratios.}
\label{fig:fig11}
\end{figure} 

The ratio $R^{*2}_{gR}/R^{*2}_{gL}$ computed here  in the thermal crossover regime, is shown by the symbols in Fig.~\ref{fig:fig11}. It increases from a value of 0.5 in a $\theta$-solvent ($z = 0$) to an asymptotic value of 0.55 in athermal solvents ($z \to \infty$). The behaviour of the ratio in the $\theta$-solvent limit has been discussed earlier in the context of Fig.~\ref{fig:Rg2_theta}~(b). As noted in section~\ref{sec:intro}, the value of the ratio reported in good solvents varies widely depending on the methodology used for its estimation. Nevertheless, several studies have reported a value of 0.56~\cite{Ragnetti85,Suzuki2011,Gartner2019}, which is close to the current prediction. Further, since,
\begin{equation} 
\left( \frac{R_{gR}}{R_{gL}} \right)^2 =  \left( \frac{R_{gR}}{R_{gR}^\theta} \right)^2 \left( \frac{R_{gR}^\theta}{R_{gL}^\theta} \right)^2 \left(  \frac{R_{gL}^\theta}{R_{gL}} \right)^2 
\label{eq:g-fac}
\end{equation}
one can use the values reported in \citet{Vlassopoulos2015} of $R_{gR}/R_{gR}^\theta =1.42$ for polystyrene rings, 
and $R_{gL}/R_{gL}^\theta= 1.36$ for linear polystyrene chains, along with $\left(R_{gR}^\theta/R_{gL}^\theta\right)^2 =0.5$, to obtain $R_{gR}^2/R_{gL}^2 = 0.55$, which agrees quantitatively with the current prediction. The values reported in  \citet{Vlassopoulos2015} were for measurements of the radius of gyration carried out in toluene as a good solvent and in cyclohexane as a $\theta$-solvent. 

It is straightforward to see from Eq.~(\ref{eq:g-fac}) that one can write the following expression for the \textit{g-factor} in the crossover regime,
\begin{equation}  
g(z)=\frac{1}{2}\frac{\alpha_{gR}^2(z)}{\alpha_{gL}^2(z)} 
\label{eq:gfit}
\end{equation}
The dashed line in Fig.~\ref{fig:fig11} displays $g(z)$ obtained from Eq.~(\ref{eq:gfit}), with the swelling functions $\alpha_{gR}$ and $\alpha_{gL}$ determined using the functional form for $f(z)$ in Eq.~(\ref{Eq:Swellratio}), and the appropriate fitting parameters given in Table~\ref{Tab:SRringlinear} for rings and linear chains. The good quality of the fit suggests the usefulness of Eq.~(\ref{eq:gfit}) in representing the \textit{g-factor} in the thermal crossover regime. Since the parameters have been determined by fitting data over a fairly limited range of $z$ values, however, the fitting function is perhaps more appropriate for determining the \textit{g-factor} by interpolation within this range of values rather than extrapolation well outside the range. 

An interesting aspect of the behaviour of the ratio $R^{*2}_{gR}/R^{*2}_{gL}$ displayed in Fig.~\ref{fig:fig11} is the rapid approach to the asymptotic athermal solvent value at fairly small values of $z$. This would suggest that measurements at even slightly elevated temperatures above the $\theta$-point in the respective solutions would saturate to the asymptotic value and it would be difficult to measure the crossover from the $\theta$-solvent to the asymptotic value. It is worth noting that  \citet{Gartner2019} have also computed a fairly constant value for the ratio close to 0.56, across a wide range of values of the quantity $\chi_{PS}$, which is used in their explicit solvent simulations as a solvent quality parameter that models the difference in the attraction strength between solvent and polymer molecules.

\begin{figure}[ptbh]
  \begin{center}
		 \hspace{-1cm}
			\resizebox{9cm}{!}{\includegraphics*[width=4cm]{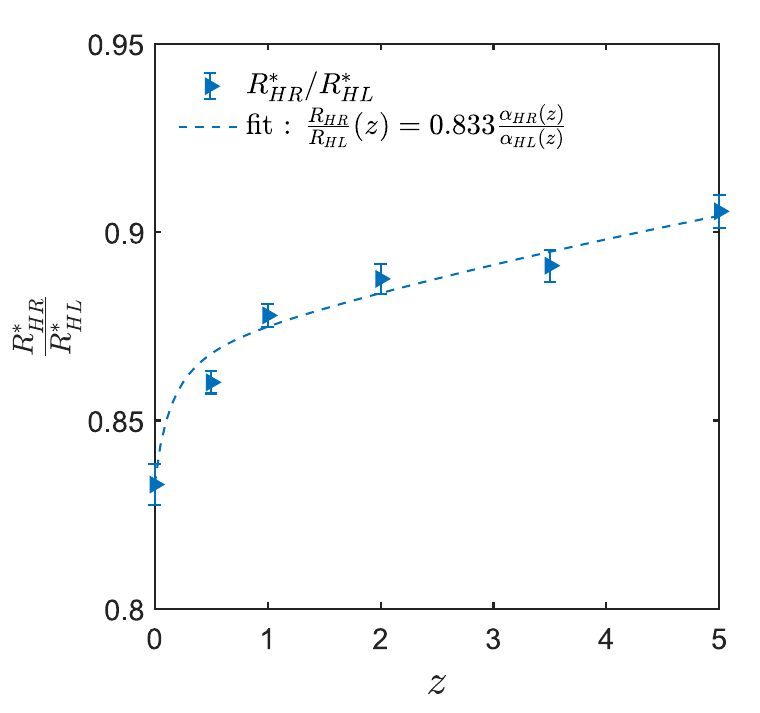}}  
	\end{center}
	\vskip-15pt
\caption{Ratio of hydrodynamic radius for ring to linear chain plotted as a function of solvent quality, $z$. The symbols represent simulation data and the dashed-line is given by the expression ${R_{HR}}/{R_{HL}}(z) = 0.833\,\left(\alpha_{HR}(z)/\alpha_{HL}(z)\right)$, where the fitting parameters in Table~\ref{Tab:SRringlinear} are used for the swelling ratios.}
\label{fig:RgRhRbyL}
\end{figure} 

As can be seen from the symbols in Fig.~\ref{fig:RgRhRbyL}, compared to the gyration radius, the ratio of the hydrodynamic radii, $R^*_{HR}/R^*_{HL}$, displays a relatively more uniform increase with the solvent quality $z$. Note that the estimation of dynamic properties requires the incorporation of fluctuating HI, which is an important aspect of the present work, and which is essential for obtaining agreement with experimental results~\cite{Prakash2019}. Under $\theta$-solvent conditions, $R_{HR}^\theta/R_{HL}^\theta = 0.833$, which  can be determined from the expression $R_{HR}^\theta/R_{HL}^\theta  = \left({U_{\text{RD}_R}^\theta}\right)^{-1} \,(R_{gR}^{\theta}/R_{gL}^{\theta}){U_{\text{RD}_L}^\theta}$, with ${U_{\text{RD}_R}^\theta}=1.167$ and ${U_{\text{RD}_L}^\theta}=1.375$, as reported earlier in section~\ref{subsec:UniRatioRh}. The ratio $R^*_{HR}/R^*_{HL}$ can be seen to be levelling off as it approaches a value around $0.9$ in the athermal limit. As in the case of the expression for the \textit{g-factor} in the crossover regime, one can derive the following expression for the hydrodynamic radii, 
\begin{align}   
\frac{R_{HR}}{R_{HL}} (z)  &=  \left( \frac{R_{HR}}{R_{HR}^\theta} \right) (z) \left( \frac{R_{HR}^\theta}{R_{HL}^\theta} \right) \left(  \frac{R_{HL}^\theta}{R_{HL}} \right) (z) \nonumber \\ &= {0.833} \left(  \frac{ \alpha_{HR}(z) }{\alpha_{HL} (z) }  \right)
\label{eq:hydradfit}
\end{align}
The dashed line in Fig.~\ref{fig:RgRhRbyL} displays the dependence of the ratio of the hydrodynamic radii on $z$ obtained from Eq.~(\ref{eq:hydradfit}), with the swelling functions $\alpha_{HR}$ and $\alpha_{HL}$ determined using the functional form for $f(z)$ in Eq.~(\ref{Eq:Swellratio}), and the appropriate fitting parameters given in Table~\ref{Tab:SRringlinear} for rings and linear chains. In this case as well, the fit is excellent, and renders Eq.~(\ref{eq:hydradfit}) a useful means of representing the ratio of hydrodynamic radii in the thermal crossover regime.

The inverse of the ratio $R_{HR}/R_{HL}$ determines the relative diffusivity of rings to linear chains, $D_R/D_L$. Its value can be computed to be around $1.11$ from the current simulations at large values of $z$. This prediction is in close agreement with the values of $1.11$ -- $1.12$ reported from scattering experiments on synthetic polymers~\cite{Higgins83,Hodgson91}. However, it deviates quite a bit from the value of 1.32 reported by \citet{Robertson2006} for DNA molecules.

\section{\label{RingConclusions} Conclusions}

Using Brownian dynamics simulations we have computed different static and dynamic properties of dilute ring polymer solutions in the thermal crossover regime between $\theta$ and athermal solvents. The use of a narrow Gaussian potential to represent excluded volume interactions enables simulations to be carried out at identical values of the solvent quality $z$ in solutions of rings and linear chains.

The universal values of the ratios involving radius of gyration $R_g$ and mean-stretch $X$ have been calculated as a function of $z$ in the limit of infinite chain length and compared with that of linear chains. Interestingly, for both linear and ring polymers the ratio $X/2R_g$ is found to be independent of the range of the excluded volume interaction potential at both finite and infinite chain lengths.

Two different methods based on the mean-squared displacement and Fixman's formula are used to evaluate the long-time diffusivity (and the hydrodynamic radius) of ring polymers. The results obtained from both the methods are in very good agreement with each other. This suggests the applicability of Fixman's formula to compute the long-time diffusion coefficient for polymeric rings. The importance of Fixman's correction for the evaluation of long-time diffusivity has been pointed out previously for linear chains, but has not been examined before for ring polymer solutions.

Universal ratios $U_{\text{RD}}$ and $X/R_H$ involving the hydrodynamic radius, $R_H$, are evaluated as a function of solvent quality, $z$, and compared with the respective values for linear polymers. In this context, we have found the fixed point value of the hydrodynamic interaction parameter to be $h^* = 0.265$ for ring polymer solutions, which leads to the quick convergence of the ratio $U_{\text{RD}}$ to its universal limit. One of the key differences in the behaviour of ring and linear polymer solutions observed in the current simulations is the dependence of $U_{\text{RD}}$ on the solvent quality $z$, with the ratio  converging asymptotically to a constant value in case of rings, while for linear chains it monotonically increases with $z$. Additionally, for polymeric rings the ratio of the mean stretch along the $x$-axis $X$ to the hydrodynamic radius $R_H$ is found to be independent of $z$, whereas the ratio depends on $z$ for linear polymers. This is consistent with the identical swelling ratios $\alpha_H$ and $\alpha_X$ observed for rings in the thermal crossover region. The significant differences highlighted here between the scaling of static and dynamic properties in ring and linear polymer solutions is an important finding of the present work that is worth validating through careful experiments.

The computation of the universal ratio $U^\theta_\text{RD}$ and the swelling curve $\alpha_H$ for rings leads to a procedure for indirectly estimating the $\theta$-temperature $\Theta_R$ for any ring polymer-solvent system, from a single measurement of the hydrodynamic radius at a temperature in the thermal crossover regime, provided the chemistry-dependent constant $k_0$ for the linear polymer-solvent system is known. 

The value for the ratio $R^{2}_{gR}/R^{2}_{gL}$ (the \textit{g-factor}) is found to increase from 0.5 in $\theta$-solvents to 0.55 in the athermal solvent limit, with the increase occurring rapidly at small values of $z$. The values in the two limits are in agreement with values reported earlier in the literature. The ratio of the hydrodynamic radii, $R_{HR}/R_{HL}$, has a more uniform increase with $z$, and its inverse $D_R/D_L$, the ratio of long-time diffusivities, approaches an asymptotic value around 1.11 in athermal solvents, close to previously reported values for synthetic polymers. Analytical expressions for both ratios are proposed in the thermal crossover regime that provide an excellent fit to the simulation data.

While the current work is focused on understanding the static and dynamic equilibrium behaviour of dilute solutions of ring polymers, the results discussed here are also relevant to the investigation of rheological and viscoelastic properties of ring polymer solutions, such as the interesting tumbling and tank-treading dynamics that have been observed recently in shear flow~\cite{Tu2020}, since even away from equilibrium, it is essential to characterise the solvent quality of a polymer-solvent system. For instance, in the case of linear polymer solutions, universal swelling curves that have been observed and computed for properties such as $R_g$, $R_H$ or the viscosity radius $R_{\eta}$, independent of the polymer-solvent chemistry~\cite{Fujita,Tominaga2002,Sharad2014,KumarRavi,Sunthar2006}, have been useful in determining the solvent quality of polymer-solvent systems from a measurement of the swelling ratios. The estimated values of $z$ have then been used to characterise systems undergoing flow~\cite{Sunthar2005b,Saadat2015,Hsiao2017,Pan2018}, and has enabled the parameter free prediction of the properties of solutions away from equilibrium and facilitated quantitative comparison with experiments~\cite{Sasmal2017,Sunthar2005,Sunthar2005b,Saadat2015,Prakash2019}. It is hoped that the current predictions of the equilibrium crossover behaviour of ring polymer solutions will prove to be similarly useful in future simulations of their rheological behaviour in various flows.

\linespread{1}\selectfont

\section*{\label{sec:acknwl}Acknowledgements}
This research was supported under Australian Research Council's Discovery Projects funding scheme (project number DP190101825). It was undertaken with the assistance of resources from the National Computational Infrastructure (NCI Australia), an NCRIS enabled capability supported by the Australian Government.

\section*{\label{sec:Appendix}Appendix: Computation of diffusivity}

\begin{figure}[tbh]
  \begin{center}
		 \hspace{-1cm}
			\resizebox{8.5cm}{!}{\includegraphics*[width=5cm]{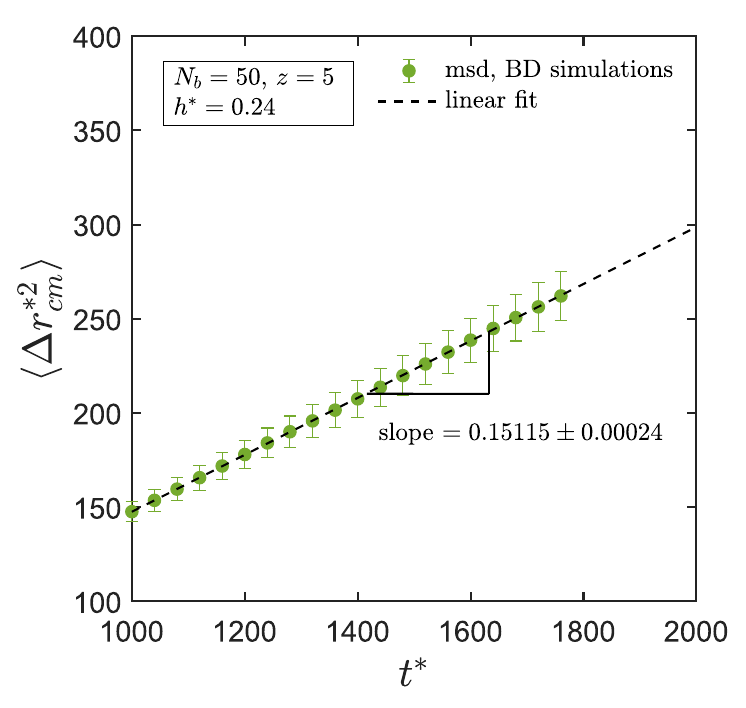}}  
	\end{center}
	\vskip-15pt
\caption{Mean-squared displacement of the non-dimensional centre of the mass as a function of non-dimensional time $t^*$ for a polymeric ring of chain length $N_b=50$, $z=5.0$ and $h^*=0.24$. The symbols are simulation data and the dashed line is a linear fit.}
\label{fig:msdNb50}
\end{figure}

\begin{figure}[tbh]
  \begin{center}
		 \hspace{-1cm}
			\resizebox{9.5cm}{!}{\includegraphics*[width=5cm]{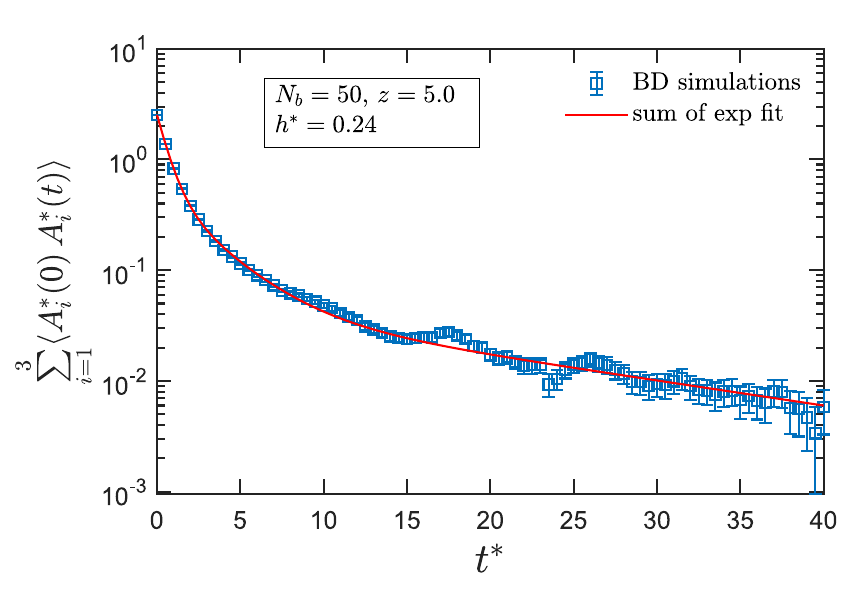}}  
	\end{center}
	\vskip-15pt
\caption{Dynamic correlation function for a polymeric ring of chain length $N_b=50$, $z=5.0$ and $h^*=0.24$. The symbols are simulation data and the solid line is a fit given by sum of exponentials.}
\label{fig:AicorrNb50}
\end{figure} 

\begin{table*}[t]
\caption{Values of $D^*_1$ and $D^*_K$ for different chain lengths ($N_b$) computed at different values of solvent quality $z$. The ratio of $D^*_1$ to $D^*_K$  as a percentage is also given to demonstrate the small difference between $D^*$ and $D^*_K$. In the last table (for $z = 5$), values of $D^*_{msd}$ calculated from the mean-squared displacement of the centre of mass of a chain is also displayed. }
\label{tab:D1Dk_data}
{\footnotesize
\begin{center}
\setlength{\tabcolsep}{5pt}
\renewcommand{\arraystretch}{1.5}
\begin{tabular}{ c  c  c  c  c  c }
\hline
 $N_b$     & $z$  & $D^*_1$  & $D^*_K$ & $({D^*_1}/{D^*_K})\times 100$  \\                
\hline
\hline
            $10$  & $0.5$ & $0.001437$ & $0.08079$ & 1.78
\\           
            $20$ & $0.5$ & $0.001469$ & $0.05575$ & 2.63
\\
            $24$ & $0.5$ & $0.00142$ & $0.05052$ & 2.81
\\
            $30$ & $0.5$ & $0.001391$ & $0.04483$ & 3.10
\\
            $40$ & $0.5$ & $0.001264$ & $0.03855$ & 3.28
\\
            $50$ & $0.5$ & $0.001194$ & $0.03417$ & 3.49
\\            
\hline
\end{tabular}
\quad
\begin{tabular}{ c  c  c  c  c  c }
\hline
 $N_b$     & $z$  & $D^*_1$  & $D^*_K$ & $({D^*_1}/{D^*_K})\times 100$   \\                
\hline
\hline
            $10$  & $1.0$ & $0.001189$ & $0.07678$ & 1.55
\\           
            $20$ & $1.0$ & $0.001173$ & $0.0521$ & 2.25
\\
            $24$ & $1.0$ & $0.001087$ & $0.04702$ & 2.31
\\
            $30$ & $1.0$ & $0.001043$ & $0.04183$ & 2.49
\\
            $40$ & $1.0$ & $0.000995$ & $0.03585$ & 2.77
\\
            $50$ & $1.0$ & $0.000942$ & $0.03179$ & 2.96
\\            
\hline
\end{tabular}
\vskip10pt
\begin{tabular}{ c  c  c  c  c  c }
\hline
 $N_b$     & $z$  & $D^*_1$  & $D^*_K$ & $({D^*_1}/{D^*_K})\times 100$  \\                
\hline
\hline
            $10$  & $2.0$ & $0.000883$ & $0.07122$ & 1.24 
\\           
            $20$ & $2.0$ & $0.000854$ & $0.04803$ & 1.78
\\
            $24$ & $2.0$ & $0.000812$ & $0.04342$ & 1.87
\\
            $30$ & $2.0$ & $0.000786$ & $0.03831$ & 2.05
\\
            $40$ & $2.0$ & $0.000745$ & $0.03275$ & 2.27
\\
            $50$ & $2.0$ & $0.000704$ & $0.02912$ & 2.42
\\            
\hline
\end{tabular}
\quad
\begin{tabular}{ c  c  c  c  c  c }
\hline
 $N_b$     & $z$  & $D^*_1$  & $D^*_K$ & $({D^*_1}/{D^*_K})\times 100$   \\                
\hline
\hline
            $10$  & $3.5$ & $0.00066$ & $0.06618$ & 1.00
\\           
            $20$ & $3.5$ & $0.000625$ & $0.04451$ & 1.40
\\
            $24$ & $3.5$ & $0.000616$ & $0.04022$ & 1.53
\\
            $30$ & $3.5$ & $0.000608$ & $0.03543$ & 1.72
\\
            $40$ & $3.5$ & $0.000571$ & $0.03035$ & 1.88
\\
            $50$ & $3.5$ & $0.000545$ & $0.02683$ & 2.03
\\            
\hline
\end{tabular}
\vskip10pt
\begin{tabular}{ c  c  c  c  c  c }
\hline
 $N_b$     & $z$  & $D^*_1$  & $D^*_K$ & $({D^*_1}/{D^*_K})\times 100$ & $D^*_{msd}$  \\                
\hline
\hline
            $24$ & $5.0$ & $0.000521$ & $0.03812$ & 1.37 & 0.03746
\\
            $30$ & $5.0$ & $0.000526$ & $0.03373$ & 1.56 & 0.03103
\\
            $40$ & $5.0$ & $0.000489$ & $0.02865$ & 1.71 & 0.02796
\\
            $50$ & $5.0$ & $0.000479$ & $0.02547$ & 1.88 & 0.02519
\\            
\hline
\end{tabular}
\end{center}
}
\end{table*}

In this paper the diffusivity of polymer chains is computed by two methods: $(i)$ from the mean-squared displacement of the center of mass of a polymer chain, $(ii)$ from Fixman's formula and the Kirkwood expression. In this section we illustrate both the methods by choosing a sample case of a polymeric ring with $N_b=50$, $z=5.0$ and $h^*=0.24$. The dimensionless diffusivity, $D^*$, is related to the mean-squared displacement of the center of mass, $\la r_\text{cm}^{*2}\ra$, by the following equation,
\begin{equation}
\la r_\text{cm}^{*2}\ra = 6\,D^*t^* 
\end{equation}
Fig.~\ref{fig:msdNb50} displays $\la r_\text{cm}^{*2}\ra$, which varies linearly with time, $t^*$. The slope of the line ($=6\,D^*$) fitting the data corresponds to the diffusivity of the polymer chain. From this method the diffusivity is calculated to be $D^* = 0.0252$.

Alternatively, diffusivity is computed using the formula $D^* = D^*_K - D^*_1$, where, $D_K^*$ is the non-dimensional Kirkwood diffusivity, defined as 
\begin{align}
D_K^* = \displaystyle\frac{1}{4N_b}+\frac{h^*\sqrt{\pi}}{4R_I^*} 
\end{align}
and $D_1^*$ is the dynamic correlation function given by Eq.~(\ref{Eq:FixmanCorr}). Fig.~\ref{fig:AicorrNb50} presents the decay of the correlation function for the system of ring chains with $N_b=50$, $z=5.0$ and $h^*=0.24$. The auto correlation function is fitted with a sum of exponentials of the form
\begin{equation}
\sum\limits_{i=1}^3 \la A^*_i(0)\,A^*_i(t)\ra = \sum\limits_{j=1}^n a_j\,\exp(b_j\,t^*)
\end{equation} 
Typically 3 to 4 exponential terms are used to fit the entire curve. Once the fitting is done, $D^*_1$ is evaluated by integrating $\sum_{i=1}^3 \la A^*_i(0)\,A^*_i(t)\ra$, analytically. For the sample case considered here, $D^*_1$ is evaluated to be $0.000479$ and $D^*_K$ is $0.02547$. This implies $D^* = 0.02499$, which is in very close agreement ($< 1.0\%$ difference) with the value of diffusivity calculated from mean-squared displacement.

Table~\ref{tab:D1Dk_data} displays the results of simulations for $D^*_1$ and $D^*_K$ at different chain lengths $N_b$ and different values of $z$. The small difference between $D^*$ and $D^*_K$ for finite chain lengths is demonstrated by the evaluation of the ratio of $D^*_1$ to $D^*_K$, which can be seen to be less than 3.5\% for all the values of $N_b$ simulated here. However, it is clear that the ratio increases with increasing values of $N_b$ at all the values of solvent quality, and as demonstrated in Fig.~\ref{fig:SwellRatio}, this leads to a significant difference in the swelling curves for $\alpha_I$ and $\alpha_H$, obtained in the long chain limit. The values of $D^*_{msd}$ obtained from simulations of the mean-squared displacement of the centre of mass of a chain is given in the last column of the table for $z = 5$, which is the only value of solvent quality for which they were determined here.

\newpage
\bibliography{JoR_Rings}
\bibliographystyle{JORnat}

\end{document}